\documentclass[a4paper,10pt]{article} \usepackage{amsmath,amssymb,amsthm}
\usepackage[english]{babel}

\newtheorem{theo}{Theorem}[section] 
\newtheorem{lemm}[theo]{Lemma} \newtheorem{prop}[theo]{Proposition}
\newtheorem{coro}[theo]{Corollary}

\newcommand{\Na}{\mathbb N}                   

\newcommand{\Za}{\mathbb Z}                   

\newcommand{\Ra}{\mathbb R}                   

\newcommand{\Ca}{\mathbb C}                   

\newcommand{\Ta}{\mathbb T}

\newcommand{\im}[1]{\mbox{Im} \ #1}

\newcommand{\scal}[1]{\langle #1 \rangle}

\newcommand{\finpreuve}{\hfill $\Box$}

\newcommand{\name}{$\underline{\qquad \qquad}$}

\newcommand{\refe}[1]{\ref{#1}} \newcommand{\reff}[1]{(\ref{#1})}

\newcommand{\e}{\epsilon}

\newcommand{\Hi}{\mathcal H}

\newcommand{\M}{\mathcal M}

\begin{document}

\author{  Jean-Marc  Bouclet\footnote{Jean-Marc.Bouclet@math.univ-lille1.fr} $
\qquad  $ Stephan  De Bi\`evre\footnote{Stephan.De-Bievre@math.univ-lille1.fr}
\\ Universit\'e  de Lille  1 \\ UMR  CNRS 8524,  \\ 59655 Villeneuve  d'Ascq }
\title{{\bf \sc  Long time  propagation and control  on scarring  for perturbed
quantized hyperbolic toral automorphisms}}


\maketitle

\begin{abstract}We show that on a suitable time scale, logarithmic in $\hbar$,
the coherent states on
the two-torus, evolved under a quantized perturbed hyperbolic toral automorphism,
equidistribute on  the torus. We then  use this result to  obtain control on
 the   possible   strong   scarring   of   eigenstates   of   the   perturbed
automorphisms by periodic orbits.  Our  main tool  is  an  adapted  Egorov theorem,
  valid  for  logarithmically long times.
\end{abstract}

\section{Introduction}
 One of  the main  results  in quantum  chaos is  the
Schnirelman  theorem. It  states  that, if  a  quantum system  has an  ergodic
classical limit, then almost all  sequences of its eigenfunctions converge, in
the classical limit,  to the Liouville measure on  the relevant energy surface
\cite{BoDB, HMR, Sc, Z1}. It is natural  to wonder if the result holds for all
sequences   (a   statement   commonly   referred  to   as   ``unique   quantum
ergodicity''). This has been proven  to be true for the (Hecke) eigenfunctions
of  the Laplace-Beltrami  operator of  a  certain class  of constant  negative
curvature surfaces \cite{li} and has been  conjectured to be true for all such
surfaces \cite{rs}.  It also has been  proven to be wrong  for quantized toral
automorphisms in \cite{fndb}. In  that case, sequences of eigenfunctions exist
 with  a semiclassical limit having  up to half of its weight supported on
a periodic orbit  of the dynamics. This phenomenon is  referred to as (strong)
scarring.  In \cite{DBBo1, fn}, it is  shown that this last result is optimal:
if a  measure is obtained as the  limit of eigenfunctions then  its pure point
component can carry at most half of its total weight.

Except for the Schnirelman theorem,  which holds in very great generality, all
cited results are proven by exploiting to various degrees special algebraic or number theoretic
properties of the  systems studied.  It is one of the  major challenges in the
field to  device proofs and  obtain results that  use only assumptions  on the
dynamical properties  of the underlying classical Hamiltonian  system, such as
ergodicity,  mixing or  exponential mixing,  the Anosov  property,  {\em etc.}
without relying on special algebraic properties.

It is argued in \cite{DBBo0, DBBo1, fn} for example,  that this will require a good
control  on the  quantum dynamics  for times  that go  to infinity  (at least)
logarithmically as the
semiclassical  parameter $\hbar$ goes  to zero:  $t\geq k_-\ln\hbar$  for some
constant $k_->0$.  It is well known that such
control is  in general hard to obtain  especially since a good  lower bound on
$k_-$ is needed. In this paper, we concentrate on
the quantized perturbed hyperbolic  automorphisms of the $2d$-torus, which are
known to be  Anosov systems classically. For those systems,  we first prove an
Egorov theorem valid for times proportional to $\ln \hbar$, with an explicit control
on the proportionality constant $k_-$ (Theorem \ref{main}).  This result is obtained by
adapting the  techniques of  \cite{BoRo1}.  We then  combine this  result with
recent sharp
estimates on the exponential mixing of the classical dynamics \cite{BKL1} to study
the long time  evolution of evolved coherent states  (Theorem \ref{cohstateevol2}), showing
that on  a sufficiently  long logarithmic time  scale, those  evolved coherent
states equidistribute on the torus. Roughly, the result is that for all $f\in C^\infty(\Ta^2)$,
\begin{eqnarray}
Q(f,t,\hbar)\equiv \left\langle  U_{\epsilon}^t
\varphi^a_{\hbar,\kappa}  ,  O  \!  p^W  (f) U_{\epsilon}^t
\varphi^a_{\hbar,\kappa}        \right\rangle_{{\mathcal
H}_{\hbar}(\kappa)}-\int_{\Ta^2} f (x) \ \mbox{d}x \rightarrow 0,
\qquad \hbar \rightarrow 0 , \label{result}
\end{eqnarray}
for times
$$
k_-\ln \hbar\leq t\leq k_+\ln \hbar, \qquad 0\leq k_- \leq k_+.
$$
Here $U_{\epsilon}$ is the unitary quantum dynamical evolution operator,
$ O  \!  p^W  (f)$ is the Weyl quantization of $f$, and
$\varphi^a_{\hbar,\kappa}$ is a coherent state at the point $a$ of the
two-torus $\Ta^2$. For detailed definitions, we refer to the following
sections.  This result generalizes results obtained in
\cite{DBBo0} for unperturbed hyperbolic automorphisms.  To prove it,
 we prove an estimate of the type
\begin{eqnarray*}
Q(f,t,\hbar)&\leq& \left| \left\langle \varphi^a_{\hbar,\kappa} ,
(U_{\epsilon}^{-t}  O  \!  p^W  (f) U_{\epsilon}^t - O \!  p^W
(f\circ \Phi_\epsilon^t )) \varphi^a_{\hbar,\kappa}
 \right\rangle_{{\mathcal H}_{\hbar}(\kappa)} \right| +\\
 &\ &\qquad\qquad \left| \left\langle \varphi^a_{\hbar,\kappa} ,
    O \!  p^W (f\circ \Phi_\epsilon^t )) \varphi^a_{\hbar,\kappa}
 \right\rangle_{{\mathcal H}_{\hbar}(\kappa)}-
\int_{\Ta^2} f (x) \ \mbox{d}x \right| \\
&\leq&\epsilon_1(\hbar e^{\gamma_q t}) + \epsilon_2(\hbar^{-1} e^{-\gamma_c t}).
\end{eqnarray*}
Here $\epsilon_1$ and $\epsilon_2$ are functions tending to zero when their
argument does. The first term comes from the error term in the  Egorov theorem, whereas the
second one involves a classical mixing rate $\gamma_c$. It is obvious that this
estimate leads to the result only if $\gamma_q< \gamma_c$. One therefore
needs $\gamma_c$  to be large (fast mixing) and $\gamma_q$ to be small.  Sharp
results on the classical mixing rates of Anosov systems
are hard to come by, but for some Anosov
maps, among which the perturbed toral automorphisms that are the subject of
this paper, such results have become available recently \cite{BKL1}.
The remaining difficulty resides therefore in controlling the exponent in the error in the
Egorov theorem. This is dealt with in the next section.

We note that, although we prove  the Egorov theorem for systems on the
$2d$-torus, we only prove the result above   in full generality for $d=1$. 
Indeed, denoting for arbitrary $d$ by $ \Gamma_{min} $ and $ \Gamma_{max} $ 
the smallest and largest 
Lyapounov exponents of the system, we prove in Section 2 that, essentially, 
$ \gamma_q
= \frac{3}{2} \Gamma_{max} $. On the other hand, the available estimates
on the classical mixing rate \cite{BKL1} yield  in our context here 
$ \gamma_c = 2 \Gamma_{min} $.
Of course, when $ d = 1 $, $ \Gamma_{max} = \Gamma_{min} $ and we have $ \gamma_q <
\gamma_c  $ as needed. This leads to \ref{result}. 
For $ d > 1 $, on the other hand, our proof of
$ \reff{result} $ still goes through, but only under an artificial 
``pinching'' condition on the Lyapounov exponents 
of the type $ 3 \Gamma_{max} < 4 \Gamma_{min} $.

As an application of the above result,
 we finally show  how to use the  information obtained on the
evolved  coherent   states  in combination with the basic strategy of
 \cite{DBBo1,DBBo1bis} to  gain   some  control  on  the   scarring  of
eigenfunctions (Theorem \ref{eigenfunctionsclassique},
 Corollary \ref{cor:eigenfunctionsclassique}). Roughly speaking, we show that
 if a sequence of
eigenfunctions  of a  quantized perturbed  hyperbolic toral  automorphism
converges to a delta measure on a finite union of periodic orbits, then 
it  must do so slowly.
An improvement on this result (basically, on how slowly)
 has been announced recently in \cite{fn2}. We don't expect this result to be
 optimal: indeed, it is expected, as in the case of unperturbed automorphisms,     
 that sequences of eigenfunctions can not concentrate
completely on periodic orbits, no matter how slowly. Proving this would
 involve controlling the quantum dynamics for longer times than we
are currently able to do.

 A result somewhat analogous to our  result on the evolution of coherent
 states was recently obtained
for the long time evolution of Lagrangian states on compact Riemannian manifolds of negative
curvature \cite{sch}. It should however be noted that such a result does not require any
control on the proportionality constant preceding $\ln \hbar$ so that no
precise control on either the mixing rate or the exponent in the error term of
the Egorov theorem are needed in that case. We suspect that in situations were such control
 can be obtained, our present strategy will allow to control both coherent
 state evolution and strong scarring.

A  related result  for  the eigenfunctions  of  Laplace-Beltrami operators  on
compact,
negatively curved Riemannian  manifolds is proven using a different strategy
in \cite{anan}:  it is shown
there  that (under a suitable technical condition that may or may not hold)
  such  eigenfunctions  can   not  concentrate  on  sets  of  small
topological entropy (and therefore on periodic orbits).

\section{Weyl quantization and Egorov Theorem}

The  purpose of  this section  is to  recall (as  compactly as  possible) some
properties of the  Weyl quantization on $  \Ta^{2d} := (\Ra / \Za  )^{2d} $ as
well as on $ \Ra^{2d} $, for $ d \geq 1 $. More specifically, we want to state
a  semi-classical version  of the  Egorov Theorem  in the  case of  $ \Ta^{2d}
$. The latter is of course well known for $ \Ra^{2d} $ but it requires a proof
for $ \Ta^{2d} $ all the more so  as we need a rather explicit version of this
theorem for the applications we have in mind in this paper.

The Weyl quantization on $ \Ra^{2d} $ can be defined as the linear map
$$  f  \in  {\mathcal  B}(\Ra^{2d})   \mapsto  O  \!  p^W  (f)  \in  {\mathcal
L}(L^2(\Ra^d)) $$  where $ O \!  p^W (f) $  is the operator (belonging  {\it a
priori}  to ${\mathcal L}({\mathcal  S}(\Ra^d),{\mathcal S}^{\prime}(\Ra^d))$)
with Schwartz kernel
$$ K_f  (q_1,q_2) = (2 \pi)^{-d} \int_{\Ra^d}  \exp ( i(q_1-q_2)\cdot p  ) \ f
\left( \frac{q_1+q_2}{2} ,  \hbar p \right) \ \mbox{d}p .  $$ Here $ {\mathcal
B}(\Ra^{2d}) $ is the set of smooth  functions $f$ on $ \Ra^{2d} $ such that $
\partial^{\gamma} f  $ is bounded  for all $  \gamma \in \Na^{2d} $,  thus the
above  integral has to  be understood  in the  sense of  oscillatory integrals
\cite{Horm3,Tayl1,Robe1,Foll1} but  it is of course a  usual Lebesgue integral
if $ f $ decays fast enough at infinity. The fact that $ O \! p^W (f) $ can be
considered  as  a  bounded operator  on  $  L^2  (\Ra^d)  $ follows  from  the
Calder\`on-Vaillancourt  theorem  \cite{Horm3,Tayl1,Robe1,Foll1} which  states
the existence of $ C > 0 $ and $ \bar{d} > 0 $ such that
\begin{eqnarray}
\left|  \left|  O  \!  p^W  (f)  \psi \right|  \right|_{L^2(\Ra^d)  }  \leq  C
\sup_{|\gamma|  \leq \bar{d}}|| \partial^{\gamma}  f ||_{L^{\infty}(\Ra^{2d})}
|| \psi ||_{L^2(\Ra^d)}, \qquad \forall \ f \in {\mathcal B}(\Ra^{2d}), \ \psi
\in {\mathcal S}(\Ra^d).
\label{CalderonVaillancourt}
\end{eqnarray}
It is  moreover well known  that $ O  \! p^W (f) $  maps the Schwartz  space $
{\mathcal S}(\Ra^d) $ continuously into itself and that $ O \! p^W(f)^* = O \!
p^W (\overline{f}) $, thus $ O \!  p^W (f) $ can be considered as a continuous
operator on  $ {\mathcal S}^{\prime}(\Ra^d) $ too.  Note also that $  O \! p^W
(f) $ is self-adjoint on $ L^2(\Ra^d) $ when $ f $ is real valued.

The Weyl quantization  on $ \Ta^{2d} $  is obtained by restricting $  O \! p^W
(f) $  to certain subspaces of $  {\mathcal S}^{\prime}(\Ra^d) $ when  $ f \in
C^{\infty}(\Ta^{2d}) $ ({\it i.e.} is $ \Za^{2d} $ periodic). The construction
is  as  follows  (see  \cite{BoDB}  for   more  details).  For  any  $  \xi  =
(\xi_q,\xi_p)  \in  \Ra^{2d}  $,   the  phase  space  translation  operator  $
U_{\hbar}(\xi) $ is defined by
$$ U_{\hbar}  (\xi) \psi (q)  = \psi (q  - \xi_q) \exp  \frac{i}{\hbar} \left(
 \xi_p  \cdot  q  -  \frac{\xi_q  \cdot \xi_p}{2}  \right),  \qquad  \psi  \in
 {\mathcal S}(\Ra^{d}) $$ and is  clearly a unitary operator on $ L^{2}(\Ra^d)
 $. One easily checks that
$$  U_{\hbar}(\xi) = O  \! p^W  (\chi_{\xi}), \qquad  \chi_{\xi} (q,p)  = \exp
\frac{i}{\hbar} (q \cdot \xi_p  - p \cdot \xi_q ) , $$  and that the following
Weyl-Heisenberg relations hold for all $ \xi , \eta \in \Ra^{2d} $
\begin{eqnarray}
U_{\hbar}(\xi)U_{\hbar}(\eta)   =  \exp  \frac{i}{2   \hbar}  \omega(\eta,\xi)
U_{\hbar}(\xi+\eta) , \label{composition}
\end{eqnarray}
with $  \omega $ the  symplectic form defined  by $ \omega (\xi,\eta)  = \xi_q
\cdot \eta_p - \xi_p \cdot \eta_q  $.  This relation shows in particular that,
if $n,m \in \Za^{2d}  $, $ U_{\hbar}(n) $ and $ U_{\hbar}(m)  $ commute if and
only if there exists $ N \in \Na $ such that
\begin{eqnarray}
2 \pi \hbar N = 1 . \label{prequant}
\end{eqnarray}
Since $ U_{\hbar}(\xi) $ acts  naturally on $ {\mathcal S}^{\prime}(\Ra) $, we
can introduce for any $ \kappa \in [0,2 \pi)^{2d} $ the space
$$ {\mathcal H}_{\hbar}(\kappa) = \{ \psi \in {\mathcal S}^{\prime}(\Ra^d) \ |
\ U_{\hbar}(n) \psi  = e^{i \omega ( \kappa  , n ) + i  \frac{n_q \cdot n_p}{2
\hbar} } \psi , \ \ \forall \ n = (n_q ,n_p ) \in \Za^{2d} \}
$$ and  it turns out that $  {\mathcal H}_{\hbar}(\kappa) $ is  of dimension $
N^d $ if $ \reff{prequant} $ holds ($0$ otherwise) with the basis
$$ \psi_r^{\kappa}(q) = N^{-  \frac{d}{2}} \sum_{k \in \Za^{2d}} e^{i \kappa_p
\cdot k} \delta_0 \left(q - k - \frac{r}{N} - \frac{\kappa_q}{2
\pi N} \right) , \qquad  r \in  \{ 0  , \cdots ,  N -  1 \}^d  .
$$ The  latter is  proven in \cite{BoDB} as  well as  the existence
of  a unique  scalar product on  each $ {\mathcal  H
}_{\hbar}(\kappa)  $ making  the  above basis  orthonormal and  $
U_{\hbar} (n/N) $ unitary for all $ n \in \Za^{2d} $. The Weyl
quantization on $ \Ta^{2d} $ is then defined by
$$   f  \in   C^{\infty}(\Ta^{2d})   \mapsto  O   \!   p^W  (f)_{|   {\mathcal
H}_{\hbar}(\kappa)} . $$ This is indeed a mapping from $ C^{\infty} (\Ta^{2d})
$ to $ {\mathcal L} ( {\mathcal H}_{\hbar}(\kappa) ) $, {\it i.e.} $ {\mathcal
H}_{\hbar}(\kappa) $ is stable under $ O  \!p^W (f) $, since one can easily check
that for any $ \Za^{2d} $ periodic function $f$
$$ O \!  p^W (f) = \sum_{n \in \Za^{2d}}  f_n U_{\hbar} (n/N) $$ if  $ f (x) =
\sum_{n } f_n e^{2i\pi  \omega (x,n)} $, $ x \in \Ra^{2d}  $. Let us emphasize
that the spaces  $ {\mathcal H}_{\hbar}(\kappa) $ are very  natural in view of
the following direct integral decomposition \cite{BoDB}
$$   L^2   (\Ra^d)    \simeq   (2   \pi)^{-2d}   \int^{\oplus}_{[0,2\pi)^{2d}}
\Hi_{\hbar}(\kappa)  \  \mbox{d}\kappa  $$  in  which the  operators  $  O  \!
p^W(f)_{|{\mathcal H}_{\hbar}(\kappa)}  $ are the fibers  of $ O \!  p^W (f) $
for this decomposition.

To streamline the discussion we will  write both quantizations on $ \Ra^{2d} $
and $  \Ta^{2d} $  under a single  form. From  now on, $  {\mathcal M}  $ will
denote  either $  \Ra^{2d} $  or $  \Ta^{2d} $.   The Weyl  quantization  on $
{\mathcal M} $ can then be defined as the map
$$ f  \in {\mathcal B}({\mathcal  M}) \rightarrow O  \! p^W (f)  \in {\mathcal
L}({\mathcal  H})  $$ where  $  {\mathcal  B} ({\mathcal  M})  $  is either  $
{\mathcal B} ( \Ra^{2d}) $ or $ C^{\infty}(\Ta^{2d}) $ and $ {\mathcal H} $ is
either  $ L^2  (\Ra^d) $  or  $ {\mathcal  H}_{\hbar}(\kappa) $  (we omit  the
$\hbar,\kappa$ dependence in the notations). In order to write $ O \! p^W(f) $
in a  unified way,  we need  to introduce the  symplectic Fourier  transform $
{\mathcal F} $ on ${\mathcal M}$ defined by
$$ {\mathcal F} f (\xi) = \int_{\M} \exp (i \omega(\xi,x)) f (x) \ \mbox{d}x
$$  where $  \xi =  (\xi_q,\xi_p) $\footnote{Throughout  this paper,  $x$ will
denote the  running point of ${\mathcal M}$  and $\xi$ the one  of $ {\mathcal
M}^* $,  unlike the usual notation  of microlocal analysis  where $(x,\xi)$ is
the running point of $\Ra^{2d}$.}  belongs to $ {\mathcal M}^* = \Ra^{2d} $ if
$ {\mathcal  M} =  \Ra^{2d} $ and  $ (2 \pi  \Za)^{2d} $  if $ {\mathcal  M} =
\Ta^{2d} $. Then the following inversion formula holds
\begin{eqnarray}
 f (x)  = \int_{{\mathcal M}^*}  \exp i \omega(x,\xi)  {\mathcal F} f  (\xi) \
\mbox{d} \nu (\xi) \label{inversionformula}
\end{eqnarray}
 with $  \mbox{d} \nu = (2 \pi)^{-2d}  \times $ the Lebesgue  measure (resp. $
\sum_{\Za^{2d}} \delta_{2 \pi n} $) if  $ {\mathcal M}^* = \Ra^{2d} $ (resp. $
(2 \pi \Za)^{2d} $) and the Weyl quantization can easily be seen to be
\begin{eqnarray}
 O \!  p^{W}(f) =  \int_{{\mathcal M}^*} U_{\hbar}  (\hbar \xi)  {\mathcal F}f
(\xi) \ \mbox{d}\nu(\xi) .  \label{quantization}
\end{eqnarray}
Of  course, when  $  {\mathcal M}  = \Ra^{2d}  $,  all the  integrals must  be
understood in the weak sense (in  $ \reff{quantization} $ we use the fact that
$ \left\langle U_{\hbar}(\hbar \xi) \psi_1 , \psi_2 \right\rangle $ belongs to
${\mathcal    S}(\Ra^{2d}_{\xi})$     if    $\psi_1,\psi_2    \in    {\mathcal
S}(\Ra^d)$).   Note  also  the   existence  of   $C,\bar{d}$  such   that,  if
$||.||_{\infty}$ denotes the $ L^{\infty} $ norm on ${\mathcal M}$,
\begin{eqnarray}
\left|  \left| O  \!  p^W (f)  \right|  \right|_{\Hi \rightarrow  \Hi} \leq  C
\sup_{|\gamma|  \leq  \bar{d} }  ||  \partial^{\gamma}  f ||_{\infty},  \qquad
\forall \ f \in {\mathcal B}({\mathcal M}) .
\label{Calderon}
\end{eqnarray}
This comes from $ \reff{CalderonVaillancourt} $ if $ {\mathcal M} = \Ra^{2d} $
and  from the unitarity  of $  U_{\hbar}(n/N) $  combined with  the elementary
estimate $ \sum_n |f_n| \leq C \sup_{|\gamma| \leq 2d+1 } || \partial^{\gamma}
f ||_{\infty} $ when $ {\mathcal M} = \Ta^{2d} $.

This  completes the  definition of  the Weyl  quantization on  $  {\mathcal M}
$. Regarding the composition of the corresponding operators, we have the
\begin{prop} \label{fcomp} There exists a bilinear map $ (f,g) \mapsto f \# g$
from $ {\mathcal B}({\mathcal M})^2 $  to $ {\mathcal B}({\mathcal M }) $ such
that
$$ O \! p^W(f)  O \! p^W (g) = O \!  p^W (f \# g). $$ The function $  f \# g $
has a full asymptotic  expansion in powers of $ \hbar $,  meaning that for all
integers $ J $
$$ f  \# g =  \sum_{ j < J}  \hbar^j f \#_j  g + \hbar^J r^{\hbar}_J  (f,g) $$
where  $  f  \#_j  g  =  \sum_{|\alpha  +  \beta|  =  j}  \Gamma(\alpha,\beta)
\partial_q^{\alpha}        \partial_p^{\beta}       f       \partial_q^{\beta}
\partial_p^{\alpha}    g   $,    with   $    \Gamma    (\alpha,\beta)^{-1}   =
(-1)^{\alpha}\alpha!\beta!(2i)^{|\alpha+\beta|}  $ and  for all  $  \gamma \in
\Na^{2d} $
\begin{eqnarray}
\left|  \left| \partial^{\gamma}  r_J^{\hbar}  (f,g) \right|  \right|_{\infty}
\leq  \frac{C_d^{J+|\gamma|}}{J!}   \sup_{|\gamma_1|  \leq  J   +  |\gamma|  +
\tilde{d}}   ||\partial^{\gamma_1}f||_{\infty}  \sup_{|\gamma_2|   \leq   J  +
|\gamma| +  \tilde{d}} ||\partial^{\gamma_2} g||_{\infty}, \qquad \  0 < \hbar
\leq 1 .
\label{remaincomp}
\end{eqnarray}
for some constants $ C_d , \tilde{d} $ depending only on $ d $.
\end{prop}

Note that  $ (. \#_j .)$  is symmetric (resp. skew  symmetric) for $  j $ even
(resp. odd) and that
\begin{eqnarray}
g \#_1 f - f  \#_1 g & = & -i (\nabla_p g \cdot  \nabla_q f - \nabla_q g \cdot
\nabla_p f) = - i \{ g,f \} . \label{Poisson}
\end{eqnarray}

\medskip

\noindent {\it Proof.} This result is well known if $ {\mathcal M}
=  \Ra^{2d} $  (see for  instance the  appendix of  \cite{BoRo1} for  a simple
proof). We  briefly sketch the proof  in the case  $ {\mathcal M} =  \Ta^{2 d}
$. Using $ \reff{composition} $ and $ \reff{quantization} $, we have
\begin{eqnarray}
 O \! p^W(f)  O \! p^W (g)  & = & \int_{{\mathcal M}^*}  \! \! \int_{{\mathcal
M}^*} e^{i \hbar \omega(\eta,\xi)/2} U_{\hbar}(\hbar(\xi + \eta)) {\mathcal F}
f (\xi)  {\mathcal F} g (\eta)  \ \mbox{d}\nu(\xi) \mbox{d}\nu(\eta) \\  & = &
\int_{{\mathcal M}^*}  U_{\hbar}(\hbar \xi) \left(  \int_{{\mathcal M}^*} e^{i
\hbar  \omega(\eta-\xi,\eta)/2} {\mathcal  F} f  (\xi -  \eta) {\mathcal  F} g
(\eta) \ \mbox{d} \nu (\eta) \right) \mbox{d}\nu(\xi).
\end{eqnarray}
Expanding $  e^{i \hbar \omega(\eta-\xi,\eta)/2}  $ by the Taylor  formula, we
get the expansion  of $ f \# g  $ with a remainder $ r_J  = r_J^{\hbar}(f,g) $
defined by its Fourier transform as follows
\begin{eqnarray}
{\mathcal F} r_J (\xi)  = \frac{(-2i)^{-J}}{(J-1)!} \int_0^1 (1-t)^{J-1} \! \!
\int_{{\mathcal M}^*} e^{i t \hbar  \omega (\eta-\xi,\eta)} \omega (\eta - \xi
, \eta)^J  {\mathcal F} f (\xi  - \eta) {\mathcal  F} g (\eta) \  \mbox{d} \nu
(\eta) \ \mbox{d} t .
\end{eqnarray}
Since  $  |   \xi^{\gamma}  {\mathcal  F}  r_J  (\xi)  |   =  |  {\mathcal  F}
{\partial}^{\widehat{\gamma}}  r_J  (\xi)  |  $,  with  $  \widehat{\gamma}  =
(\gamma_p,\gamma_q) $ if $  \gamma = (\gamma_q,\gamma_p) $ ($\gamma_q,\gamma_p
\in \Na^{d}$), we have to consider
$$ \xi^{\gamma}  \omega (\eta -  \xi , \eta)^J  = \sum_{\gamma_1 +  \gamma_2 =
\gamma}   \sum_{|\beta|  =   J}   (-1)^{|\beta_q  |+|\gamma_1|}   \frac{\gamma
!}{\gamma_1 ! \gamma_2  !} \frac{J!}{\beta !} (\eta -  \xi)^{\beta + \gamma_1}
\eta^{\widehat{ \beta  } +  \gamma_2} $$ where  $ \beta =  (\beta_q,\beta_p) $
with  $  \beta_q  ,  \beta_p  \in   \Na^d  $.  The  sum  contains  at  most  $
C_d^{J+|\gamma|} $ terms and since
$$  \frac{J !}{\beta  !} \leq  (2d)^J ,  \qquad \frac{  \gamma !}{  \gamma_1 !
\gamma_2 !} \leq  2^{| \gamma |} $$ we conclude that  $ \reff{remaincomp} $ is
now a simple consequence of the fact that
$$  \int_{{\mathcal  M}^*}  \left| \xi^{\widehat{\alpha}}  {\mathcal  F}f(\xi)
\right|   \  \mbox{d}\nu(\xi)   =  \int_{{\mathcal   M}^*}   \left|  {\mathcal
F}\partial^{\alpha}f   (\xi)  \right|   \  \mbox{d}\nu(\xi)\leq   C_d  \sup_{|
\alpha_1| \leq 2d+1} || \partial^{ \alpha_1 + \alpha}f||_{\infty} . $$ We omit
the details. \finpreuve

\medskip

\noindent {\it Remark.} The above proof can be repeated {\it
verbatim}  if $  {\mathcal M}  = \Ra^{2d}  $ and  ${\mathcal  B}(\Ra^{2d})$ is
replaced by $ {\mathcal S}(\Ra^{2d}) $.

\medskip

We now present a unified version  of Egorov Theorem, that is the semiclassical
analysis of  $ e^{ i t O  \!p^W(g)/ \hbar} O \!  p^W (f) e^{-it O  \! p^W(g) /
\hbar }  $ for  $ f,g \in  {\mathcal B}(\M) $,  with $  g $ real  valued. This
result is well known for  $ \M = \Ra^{2d} $ \cite{Egor1,Tayl1,Horm3,Robe1} and
the purpose  of what follows  is essentially to  prove a similar result  for $
{\mathcal M}  = \Ta^{2d}  $, with  an explicit remainder  term. The  result is
based  on the following  simple remark:  if $  A $  is a  bounded self-adjoint
operator and $B(t) $ is a strongly $C^1$ family of bounded operators, then
\begin{eqnarray}
e^{itA/\hbar}B(0)e^{-itA/\hbar}   -   B   (t)   =   \frac{i}{\hbar}   \int_0^t
e^{i(t-s)A/\hbar}   \left(  \hbar  i   \frac{d}{ds}B(s)  +   [A,B(s)]  \right)
e^{-i(t-s)A/\hbar}. \label{Duhamel}
\end{eqnarray}
 We shall use this formula with $ A = O \! p^W (g) $ and $ B (t) $ of the form
$$  B(t) = \sum_{j<J}  \hbar^j O  \! p^W  (f_j(t)) $$  with $  f_0(t), \cdots,
f_{J-1}(t) \in {\mathcal  B}({\mathcal M}) $ such that $  \sum_{j < J} \hbar^j
f_j (0) = f$ ({\it i.e.}  $B(0) = O \! p^W (f) $) and
\begin{eqnarray}
 \hbar i \frac{d}{ds} B(s) + [A,B(s)] = {\mathcal O}(\hbar^{J+1})
\label{equations}
\end{eqnarray}
 where $  {\mathcal O}(h^{J+1}) $ is to  be understood the operator  norm on $
\Hi $. Expanding $ [A,B(s)] $ in powers  of $ \hbar $ by means of Proposition $
\refe{fcomp} $, $ \reff{equations} $  leads to the following conditions on the
functions $ f_{j}(s) $
\begin{eqnarray}
\partial_s f_0  - \{ g  , f_0 \}  & = &  0 , \qquad  \qquad \qquad \qquad  \ \
f_0(0) = f,  \\ \partial_s f_j - \{  g , f_j \} &  = & 2 i \sum_{l+k  = j+1} g
\#_k f_l, \qquad f_j (0) = 0 \ \ \mbox{for} \ j \geq 1
\end{eqnarray}
where, in the  last sum, $ 3 \leq  k \leq J - l  $ is odd and $ l  \leq J-1 $,
which implies actually that $ l \leq j-2 $. This system is thus triangular and
can  be solved using  the Hamiltonian  flow $  \phi^s $  of $  g $,  since the
solution of $ \partial_s a - \{ g, a \}  = b $ with $ a_{s=0} = a_0 $ is given
by
$$  a  (s,x)  =  a_0  (\phi^s(x))  + \int_0^s  b  (\tau,  \phi^{s-\tau}(x))  \
 \mbox{d}\tau, \qquad  x \in {\mathcal  M}. $$ Note  that if $ {\mathcal  M} =
 \Ta^{2d} $ and $ g $ is identified with a $ \Za^{2d} $ periodic function on $
 \Ra^{2d} $, the associated Hamiltonian  flow $ \tilde{\phi}^s $ on $ \Ra^{2d}
 $  is   easily  seen  to   satisfy  the  identity  $   \tilde{\phi}^s(x+n)  =
 \tilde{\phi}^s(x) +  n $  for all $  x \in  \Ra^{2d} $ and  $ n  \in \Za^{2d}
 $.  This shows  that  the formulas  for  the $f_j(s)  $ are  the  same for  $
 {\mathcal M} = \Ra^{2d}  $ and $ \Ta^{2d} $, if $f$ and  $g$ are $ \Za^{2d} $
 periodic.

Let us now define the linear operators $L_j^s$ on ${\mathcal B}({\mathcal M})$
by $ L_j^s f := f_j (s) $. We have
\begin{eqnarray}
 L_0^s  f =  f \circ  \phi^s,  \qquad L_j^s  \equiv 0  \  \ \mbox{for}  \ j  \
\mbox{odd} \label{Lzero}
\end{eqnarray}
 the latter  being a consequence of  the (skew) symmetry  of $\#_k$ for $  k $
 (odd) even.  For $ j \geq 2 $ even, an induction shows that
\begin{eqnarray}
L_j^s  = \frac{1}{(2i)^j}  \sum_{k=1}^{j/2} \sum_{m_1  + \cdots  + m_k  = j/2}
\sum_{|\alpha_1  + \beta_1|=1+  2m_1}  \cdots \sum_{|\alpha_k+\beta_k|=1+2m_k}
\qquad  \qquad  \nonumber  \\  \int_0^s  \cdots  \int_0^{s_{k-1}}  L_0^{s-s_1}
M^{\alpha_1,\beta_1}  L_0^{s_1-s_2}  \cdots  M^{\alpha_k,\beta_k} L_0^{s_k}  \
\mbox{d}s_k \cdots \mbox{d}s_1 \label{Ljs}
\end{eqnarray}
where $ m_1 \geq  1, \cdots, m_k \geq 1 $ in the  sum and $ M^{\alpha,\beta} $
is the differential operator
\begin{eqnarray}
 M^{\alpha,\beta}  = \frac{(-1)^{|\alpha|}}{\alpha!\beta!}  \partial_q^{\beta}
\partial_p^{\alpha}g           \partial_q^{\alpha}          \partial_p^{\beta}
. \label{operateurdifferentiel}
\end{eqnarray}
Taking the remainders into account, one gets the following result:
\begin{theo}[Egorov theorem] \label{Egorov} For all $f,g \in {\mathcal B}({\mathcal M})
$ with $ g$ real valued and all $J \geq 1 $ we have
$$ e^{it O \! p^W(g)/\hbar } O \! p^W (f) e^{-it O \! p^W (g)/\hbar} = \sum_{j
< J} \hbar^j O \! p^W (L_j^t f) + \hbar^J R_J^t(f,\hbar)
$$ where the operator $ R_J^t(f,\hbar) $ has the following explicit form
$$   i   \int_0^t  e^{i(t-s)O   \!   p^W   (g)/\hbar}   \left(  \sum_{l<J}   O
\!p^W(r^{\hbar}_{J-l+1}(g,L_l^s f)) -  O \! p^W (r_{J-l+1}^{\hbar}(L^s_l f,g))
\right) e^{-i(t-s) O \!  p^W(g)/\hbar} \ \emph{d}s . $$
\end{theo}

\medskip

Note that estimates on $ || R_J^t(f,\hbar)||_{\Hi \rightarrow \Hi } $ can then
be derived  from $ \reff{Calderon} $,  $ \reff{remaincomp} $  and estimates on
the derivatives  of $ L_j^s  f $.  This will be  extensively used in  the next
section.

\section{Perturbations of quantized hyperbolic maps} \label{Cat}
\setcounter{equation}{0}  In  this section,  we  address  the  problem of  the
semi-classical approximation of $ U_{\epsilon}^{-t}  O \!p^W (f) U_{\e}^t $ as
$ \hbar \downarrow 0 $ in the  Ehrenfest time limit $ t \approx |\ln \hbar| $,
when $ U_{\e} $ is a unitary operator on $ {\mathcal H} $ of the form
$$ U_{\epsilon} = e^{- i \epsilon O \! p^W  (g)/ \hbar } M (A) $$ with $ M (A)
$ the quantization  of a symplectic matrix  with integer entries $ A  \in S \!
p(d,\Za)  $. We  refer  to \cite{BoDB,DBBo0,DBBo1}  and  \cite{Foll1} for  the
definition  of $M(A)$  by mean  of the  metaplectic representation  of $  S \!
p(d,\Ra) $ and only quote the properties  that we need. The operator $ M (A) $
is defined, up  to a phase, as the unique operator  on $ {\mathcal S}^{\prime}
(\Ra^d) $ such that
\begin{eqnarray}
 M(A)^{-1} O \! p^W (f) M (A) = O  \! p^W (f \circ A) , \qquad \forall \ f \in
{\mathcal B}(\Ra^{2d}) . \label{Egorovexact}
\end{eqnarray}
 If $ {\mathcal M} = \Ra^{2d} $,  $M(A)$ is unitary on $ L^{2}(\Ra^{d}) $, but
if  $ {\mathcal M}  = \Ta^{2d}  $ and  $ {\mathcal  H} =  {\mathcal H}_{\hbar}
(\kappa) $ one has  to choose special values of $ \kappa $  to ensure that $ M
(A) $ maps $ {\mathcal H}_{\hbar}(\kappa)  $ into itself, in which case $M(A)$
is unitary  (see \cite{BoDB} for more  details); from now on,  we shall assume
that  such a  choice, which  depends on  $  \hbar $,  has been  made. Then,  $
\reff{Egorovexact} $ holds on $ {\mathcal M} = \Ra^{2d} $ and $ \Ta^{2d} $ and
this is often expressed by saying that for linear evolutions `Egorov is
exact', meaning there is no remainder term.

Let  us  now describe  the  results  of this  section.  We  will  denote by  $
\phi^{\epsilon} $ the  Hamiltonian flow associated to a fixed  real valued $ g
\in  {\mathcal  B}({\mathcal   M})  $  and  consider  the   discrete  group  $
(\Phi_{\epsilon}^t)_{t  \in \Za}  $  of symplectomorphisms  on ${\mathcal  M}$
defined by
\begin{eqnarray}
 \Phi_{\epsilon} = \phi^{\e} \circ A .
\end{eqnarray}
Then, by setting $$ \tilde{L}_j f = (L^{\e}_j f) \circ A $$ with the notations
of $\reff{Lzero}$ and $\reff{Ljs}$, we can consider the functions
$$ {\mathcal  L}_{0}^t f =  f \circ \Phi_{\e}^t,  \qquad {\mathcal L}^t_j  f =
\sum_{l_1  + \cdots +  l_t =  j} \tilde{L}_{l_1}  \cdots \tilde{L}_{l_t}  f $$
defined for $  j \geq 1 , t \geq 0  $ integers and $ f \in  {\mathcal B}( \M )
$. Note that they depend on $  \e $ but we omit this dependence for notational
convenience.  Note also that $ {\mathcal L}_{0}^t = (\tilde{L}_0)^t $ and that
${\mathcal L}^t_j \equiv 0 $ if $ j $ is odd. Our goal is to show that
$$ U_{\epsilon}^{-t} O \!p^W (f) U_{\e}^t  \sim \sum h^j O \! p^{W} ({\mathcal
L}^t_j f) , \qquad \hbar \downarrow 0 , $$ in a scale of times $ t $ described
in terms of exponents $ \Gamma_A , \Gamma_g $ that we now define.

\smallskip

For the sake of simplicity, we shall  assume that $ A $ is diagonalizable over
$  \Ra $,  meaning that  there exists  an invertible  matrix $  P $  with real
entries such  that $ A  = P^{-1}D P $  with $ D  $ diagonal. Note that  such a
condition is  of course satisfied if  $ A $  is symmetric, {\it e.g.}  the cat
map.   At  the end  of  the  section, we  explain  how  to  cope with  general
symplectic matrices $ A \in S \! p  (d,\Za) $. Let us define $ \Gamma_A \geq 0
$ by
$$ e^{\Gamma_A} =  \sup_{\sigma(A)}|\lambda| . $$ Of course,  this quantity is
well defined  for any invertible matrix $  A $ with real  or complex spectrum.
For $ z = (z_1,\cdots,z_{2d}) \in \Ca^{2d}  $, we denote by $|z| := (|z_1|^2 +
\cdots |z_{2d}|^2 )^{1/2}$ its standard hermitian  norm and set $ ||z||_P := |
P z | $. The interest of the norm $||.||_P $ is that we have
\begin{eqnarray}
||A  z  ||_P  \leq  e^{\Gamma_A}||z||_{P},   \qquad  ||  \im  A  z  ||_P  \leq
e^{\Gamma_A} ||\im z||_P \qquad \forall \ z \in \Ca^{2d} ,
\label{dilateA}
\end{eqnarray}
which we shall use extensively in the sequel.

Then,   inspired  by   \cite{Trev1,BoRo1},   we  define   the   open  sets   $
\Omega_{\delta} \subset \Ca^{2d} $ for $ \delta > 0 $ by
$$ \Omega_{\delta} = \{ z \in \Ca^{2d} \ | \ ||\im z||_P < \delta \} $$ and we
consider  the family  of norms  $||.||_{\tau,\delta}$ defined  for $  \tau \in
(0,1) $ by
$$  ||f ||_{\tau,\delta}  =  \sup_{z \in  \Omega_{\tau  \delta}}|f(z)| $$  for
functions $ f $ which are bounded  and analytic on $ \Omega_{\delta} $. We can
now set
$$ \Gamma_g = \sup_{z \in \Omega_{\delta}}||| {\mathcal J} \nabla^2 g(z)|||_P,
\qquad {\mathcal J} =
\begin{pmatrix}
  0 & I \\ -I & 0
\end{pmatrix} $$
with $ \nabla^2 g $ the Hessian matrix of $g$ and $ ||| B |||_P := \sup_{z \ne
0} || B z ||_P / ||z||_P $ for $ B \in M_{2d}(\Ca)$.  Note that $ \Gamma_g \ne
0 $ unless $g$ is constant which is a trivial situation. We then define
$$ \Gamma_{\epsilon} = \Gamma_A + \epsilon  \Gamma_g $$ and our main result is
the following:
\begin{theo} \label{main} Assume that $f,g \in {\mathcal B}({\mathcal M}) $, with $ g$ real valued,
 have bounded and analytic extensions to $ \Omega_{\delta} $ for some $ \delta
 > 0 $.  Then, for all $0 < \nu < 2  $, there exists $ J_0 > 0 $ such that for
 all $ J > J_0 $
\begin{eqnarray}
U_{\e}^{-t} O  \! p^W  (f) U_{\e}^t  = \sum_{j <  J} \hbar^j  O \!  p^W \left(
{\mathcal L}_j^t f \right) + \hbar^J \varrho_J^t(f,\e,\hbar) \label{expansion}
\end{eqnarray}
with a remainder such that, for all $ 0 \leq \e \leq 1 $,
\begin{eqnarray}
\left|  \left| \hbar^J  \varrho_J^t(f,\e,\hbar) \right|  \right|_{{\mathcal H}
\rightarrow {\mathcal H}} \rightarrow 0 \qquad \mbox{as} \ \ \hbar \rightarrow
0 \qquad \mbox{if} \  \ \ 0 \leq t \leq \frac{2  - \nu}{3 \Gamma_{\epsilon}} |
\ln \hbar | .
                        \label{nontrivial}
\end{eqnarray}
\end{theo}

The reader may wonder what  $ \reff{nontrivial} $ means if $ \Gamma_{\epsilon}
=  0 $.  In  such a  case $  \Gamma_g  = 0  $  thus $  g  $ is  constant so  $
\reff{expansion} $ becomes $ U_{\epsilon}^{-t} O \! p^W (f) U_{\epsilon}^t = O
\! p^W (f \circ A^t) $ by $  \reff{Egorovexact} $ which holds for all $ t \geq
0 $.   In Section 3,  we will  anyway be interested  in the situation  where $
\Gamma_A > 0 $ and $ \epsilon $ is small so that $ \Gamma_{\epsilon} > 0 $.

We also  emphasize that the analyticity  assumption is imposed by  our need to
control  high  order  derivatives of  $f$  and  $g$  in  order to  estimate  $
\varrho_J^t(f,\e,\hbar) $. Similarly to  \cite{BoRo1}, we could probably relax
such a  condition by considering  quasi-analytic functions ({\it  e.g.} Gevrey
functions) which would allow us to consider compactly supported $f$.

\smallskip

The rest of this section is now  devoted to the proof of Theorem $ \refe{main}
$.  The principle is  rather simple  and is  the following:  a straightforward
application of Theorem $ \refe{Egorov} $ shows that
$$ U_{\e}^{-1} O \! p^W (f) U_{\e} = \sum_{j < J} \hbar^j O \! p^W(\tilde{L}_j
f) + \hbar^J  M(A)^{-1} R_J^{\e}(f,\hbar) M (A)  , \qquad \forall \ J  > 0 ,$$
hence an induction on $ t \geq 1 $ shows that $ \reff{expansion} $ holds with
\begin{eqnarray}
 \varrho_J^t(f,\e,\hbar)  = \sum_{j=J}^{tJ}  \hbar^{j-J} O  \!  p^W ({\mathcal
K}^t_{j,J}  f)  + \sum_{s  =1}^t  U_{\e}^{s-t}  M (A)^{-1}  R_J^{\e}({\mathcal
E}_J^{s-1}f,\hbar) M(A) U_{\e}^{t-s} \label{restecomposition}
\end{eqnarray}
with the operators $ {\mathcal K}^t_{j,J} $ and $ {\mathcal E}_J^{t} $ defined
by
\begin{eqnarray}
 {\mathcal K}^t_{j,J} = \sum_{l_1 + \cdots + l_t  = j \atop l_1 < J , \cdots ,
l_t < J} \tilde{L}_{l_t} \cdots  \tilde{L}_{l_1} , \qquad {\mathcal E}_J^{t} =
\sum_{l_1  <   J}  \cdots   \sum_{l_t  <  J}   \hbar^{l_1  +  \cdots   +  l_t}
\tilde{L}_{l_t} \cdots \tilde{L}_{l_1} . \label{KEjJt}
\end{eqnarray}
 Note that $ {\mathcal K}^t_{j,J} $ depends on both $ j $ and $ J $ unless $ j
< J $ in which case $ {\mathcal K}^t_{j,J} = {\mathcal L}^t_j $.  Note moreover
that $ {\mathcal E}_J^{t}  $ depends on $ \hbar $ and  that we set $ {\mathcal
K}_{j,J}^1 = 0 $, $ {\mathcal E}_J^0 = \mbox{id} $.

Thus  $\reff{Calderon}$   reduces  the  proof  of  Theorem   $  \refe{main}  $
essentially  to   estimate  the   derivatives  of  $   \tilde{L}_{l_t}  \cdots
\tilde{L}_{l_1} f  $. To that end, we  shall use the following  extension of a
lemma of \cite{BoRo1}.
\begin{lemm} \label{lemmecopie} There exists a constant $ C_P $ depending only on $ P
$ such that, if
\begin{eqnarray}
||f||_{\tau,\delta}  \leq  M  \left(  \frac{C_P}{1-\tau}  \right)^{a},  \qquad
\forall \ 0<\tau <1 \label{diffborne}
\end{eqnarray}
for some $ M , a \geq 0 $, then for all $ \gamma $ we have
\begin{eqnarray}
 || \partial^{\gamma}   f||_{\tau,\delta}   \leq   M   (a+|\gamma|)\cdots(a+1)
\delta^{-|\gamma|}   \left(  \frac{C_P}{1-\tau}   \right)^{a+|\gamma|}  \qquad
\forall \ 0<\tau <1 . \nonumber
\end{eqnarray}
\end{lemm}

\noindent {\it Proof.} In \cite{BoRo1}, the authors show that the
result holds  with $ P =  I $ and  $ C_P = e  $. Our lemma follows  from their
result applied to $ f \circ P^{-1} $.
 \finpreuve

\medskip In order to estimate $ f  \circ \Phi_{\e} $ we will capitalize on two
facts: on one hand, $ \reff{dilateA} $ implies that
\begin{eqnarray}
 || f \circ A ||_{\tau,e^{-\Gamma_A} \delta} \leq ||f||_{\tau,\delta}
    \label{flowA}
\end{eqnarray}
 and on the other hand we have, for any $ 0 \leq s \leq t $,
$$  ||  \im  z  ||_P  \leq  \tau  \delta  e^{-t\Gamma_g}  \Rightarrow  ||  \im
\phi^{t-s}(z) ||_P \leq \tau \delta e^{-s \Gamma_g} .
$$ The latter is actually shown in $  \cite{BoRo1} $ only for $ P= I $ but the
very same method easily leads to this estimate.
We therefore omit the proof and rather emphasize that it implies that
\begin{eqnarray}
 || f   \circ   \phi^t  ||_{\tau,   \delta   e^{-t   \Gamma_g}}   \leq  ||   f
    ||_{\tau,\delta}
\label{flowP}
\end{eqnarray}
 which leads to the
\begin{lemm} \label{lemmecopie2} Assume that  $ C_g > 0 $ is such that
$ | \partial^{\gamma}  g (z)| \leq \gamma  ! C_g^{|\gamma|} $ for all  $ z \in
\Omega_{\delta}  $  and all  $  |\gamma|  \geq 1  $.  Assume  moreover that  $
\reff{diffborne} $ holds. Then for all $ j \geq  2 $ even and all $ s \geq 0 $
real, we have
$$ ||L_j^s f ||_{\tau, \delta e^{-s\Gamma_g}} \leq M \left( \frac{C_P}{1-\tau}
\right)^{a + \frac{3j}{2}} e^{\frac{3j}{2}s  \Gamma_g} (a+1) \cdots \left( a +
\frac{3j}{2}  \right) (4d C_g/\delta)^{3j/2}\sum_{k=1}^{j/2}  \frac{s^k}{k!} ,
$$ for all $ \tau \in (0,1) $.
\end{lemm}

\medskip

\noindent {\it Proof.} We first note that, by an easy induction on
$ k \geq  0 $, the following  result hods: if $ s_0  , \cdots , s_k  $ are non
negative  real   numbers  such  that   $s_0  +  \cdots   +  s_k  =  s   $  and
$\alpha_1,\beta_1, \cdots , \alpha_k,\beta_k$  are non zero multi-indices such
that $|\alpha_1 + \beta_1| + \cdots + |\alpha_k + \beta_k | = n$, then for all
$ \tau \in (0,1) $
$$  || L_0^{s_k}  M^{\alpha_k,\beta_k}  \cdots L_0^{s_1}  M^{\alpha_1,\beta_1}
L_0^{s_0}  f ||_{\tau,  \delta e^{-s\Gamma_g}}  \leq M  \left(  \frac{C_P}{1 -
\tau} \right)^{a+n} C_g^{n}  \delta^{- n} e^{s n \Gamma_g}  (a+1) \cdots (a+n)
.  $$ This  follows from  Lemma  $ \refe{lemmecopie}  $ and  $ \reff{flowP}  $
(recall that $M^{\alpha,\beta}$ is defined by $\reff{operateurdifferentiel}$).
The lemma  is then  a consequence of  $ \reff{Ljs}  $ combined with  the above
estimate, the fact that
\begin{eqnarray*}
 \# \{ (\alpha,\beta) \in \Na^{2d}  \ | \ |\alpha + \beta| = 1 +  2 m \} & = &
\frac{(2d+2m)!}{(2d-1)!(2m+1)!}  \leq (2d)^{1+2m} ,  \\ \#  \{ (m_1,  \cdots ,
m_k) \in  \Na^k \ |  \ m_1 +  \cdots + m_k  = j /  2 \} &  = & \frac{(k-1  + j
/2)!}{(k-1)! (j/2)!}  \leq 2^{k-1+\frac{j}{2}},
\end{eqnarray*}
and  the fact  that $  \int_0^s \cdots  \int_0^{s_{k-1}} \  \mbox{d}s_k \cdots
\mbox{d}s_1 = s^k / k! $. \finpreuve

\medskip

We can now state the main ingredient of the proof of Theorem $ \refe{main} $.
\begin{prop} \label{almost} With the same assumptions as in Lemma
$ \refe{lemmecopie2} $, we have: for all $ j \geq 2 $ and all integers $ l_1 ,
\cdots , l_t  $ such that $l_1 + \cdots +  l_t = j$, we have for  all $ \e \in
[0,1] $
\begin{eqnarray}
|| \tilde{L}_{l_t}   \cdots   \tilde{L}_{l_1}   f   ||_{\tau,   \delta   e^{-t
\Gamma_{\epsilon}   }}   \leq    M   \left(   \frac{C_P}{1-\tau}   \right)^{a+
\frac{3j}{2}}  e^{\frac{3t}{2}  j \Gamma_{\epsilon}}  (a+1)  \cdots \left(a  +
\frac{3j}{2} \right) (4 d C_g e^{\e} / \delta )^{3j/2} \nonumber
\end{eqnarray}
provided that $ \reff{diffborne} $ holds. In addition,
 if $
|f(z)| \leq M $ on $ \Omega_{\delta} $, there exists  a constant $ K $ such that,
for all $ t \geq 1 $, all $\gamma$ and all $ \e \in [0,1] $
\begin{eqnarray}
 ||\partial^{\gamma} {\mathcal K}^t_{j,J} f ||_{\infty} \leq M t^j K^{(1+\e)j}
 (|\gamma|+3j/2)!  e^{ t \Gamma_{\epsilon}  (|\gamma|+3j/2) }, \qquad 0 \leq j
 \leq t J. \label{suiteCalderon}
\end{eqnarray}
\end{prop}

\noindent {\it Proof.} Recall that we can assume that $ j $ is
even.  We obtain the first statement by induction on $ t \geq 1 $ using lemma $
\refe{lemmecopie2} $ with $ s = \e $ and $ \reff{flowA} $ which we use through
$ ||  f \circ  A ||_{\tau,  e^{- 3  \Gamma_A / 2}  \delta} \leq  || f  \circ A
||_{\tau, e^{-  \Gamma_A } \delta} \leq ||f||_{\tau,\delta}  $. This, together
with $ \reff{KEjJt} $ then yields the second statement
since $ \# \{ l_1  + \cdots + l_t = j \} \leq t^j $.\finpreuve

\bigskip

\noindent {\bf Proof of Theorem $ {\bf \refe{main}} $.}  We first  estimate $ ||
\sum_{j=J}^{tJ} h^{j}  O \! p^w  ({\mathcal K }^t_{j,J}  f)||_{\Hi \rightarrow
\Hi} $. Using $ \reff{Calderon} $, $ \reff{suiteCalderon} $ allows to estimate
$ \ln  \left( ||  h^{j} O \!  p^w ({\mathcal K}^t_{j,J}  f)||_{\Hi \rightarrow
\Hi} \right) $ from above by
$$  j \left\{ \ln  \hbar +  t \left(  \frac{3}{2} +  \frac{\bar{d}}{J} \right)
 \Gamma_{\epsilon}  + \ln  ( K^2t  ) +  \left(\frac{3}{2}  + \frac{\bar{d}}{J}
 \right) \ln (\bar{d}+ 3j/2) \right\} + \ln C M
$$ using  the fact  that $ 1/j  \leq 1 /  J $ and  that $  \e \in [0,1]  $. By
choosing $ J $ large enough we can assume that
$$ \frac{3}{2}  + \frac{\bar{d}}{J} \leq \frac{3 }{2  - \nu/2} . $$  Since $ j
 \leq   J  t   $,  the   term   $  \ln   (\bar{d}+3j/2)  $   is  $   {\mathcal
 O}(\ln|\ln|\hbar||) $ as $ \hbar \downarrow  0 $ thus we get the existence of
 a new constant $ C $ such that for all $ \nu \in (0,2)$, $ \hbar \in (0,1] $,
 $ \e \in [0,1] $, $ 1 \leq t \leq (2 - \nu)/3 \Gamma_{\epsilon} $ and $ j \in
 [J,tJ] $
\begin{eqnarray}
 || h^{j} O  \! p^w ({\mathcal K  }^t_{j,J} f)||_{\Hi \rightarrow  \Hi} \leq C
\hbar^{\nu  j /2} \left(  \ln (C  + |\ln  \hbar|) \right)^{Cj}  \leq \tilde{C}
\hbar^{ \nu j / 4} .
\end{eqnarray}
Since $  \sum_{J \leq j \leq  t J} $ contains  $ {\mathcal O}(  |\ln \hbar|) $
terms, we see that $ ||  \sum_{j=J}^{tJ} h^{j} O \! p^w ({\mathcal K }^t_{j,J}
f)||_{\Hi \rightarrow \Hi} \rightarrow 0 $.

Now the  norm of second  term of $  \reff{restecomposition} $ multiplied  by $
\hbar^J $ can be estimated by
\begin{eqnarray}
 t  J h^J  \sup_{0 \leq  \tau \leq  t-1 \atop  s \in  [0,\e], \  l <  J}  || O
\!p^W(r^{\hbar}_{J-l+1}(g,L_l^s {\mathcal  E}_J^{\tau} f)) ||_{\Hi \rightarrow
\Hi} + ||O \! p^W (r_{J+l-1}^{\hbar}(L^s_l {\mathcal E}_J^{\tau} f,g)) ||_{\Hi
\rightarrow \Hi}
 \label{deuxieme}
\end{eqnarray}
with  the notations  of Theorem  $ \refe{Egorov}  $. We  proceed as  before to
estimate  $  L_l^s  {\mathcal E}_J^{\tau}  f  $  and  we obtain  the  theorem.
\finpreuve

\bigskip

Let us now briefly describe how to prove  such results for a general $ A \in S
\! p (d,  \Za) $ with $ \Gamma_A >  0 $. We claim that, in  this case, we have
the following result: for any $  \tilde{\Gamma}_A > \Gamma_A $ there exists an
invertible matrix $ P $ with real entries such that
\begin{eqnarray}
 ||| P^{-1} A P ||| \leq e^{\tilde{\Gamma}_A }
\label{nilpotent}
\end{eqnarray}
where $ |||.|||  $ is the matrix  norm associated to the hermitian  norm $ |.|
$ on $\Ra^2$. We can prove this statement as follows. Assume first that the spectrum of $
A $ is real and let us choose a basis $(e_1,\cdots,e_{2d})$ of $ \Ra^{2d} $ in
which  $ A  $  is in  Jordan  normal form.  If  $ (e_j,  \cdots  , e_{j+p})  $
corresponds to a Jordan block
$$ J (\lambda) = \left( \begin{array}{ccccc} \lambda & 1 & 0 & \cdots & 0 \\ 0
& \lambda &  1 & \ddots & \vdots \\  \vdots & \ddots & \ddots &  \ddots & 0 \\
\vdots & & \ddots &\ddots & 1 \\ 0 & \cdots & \cdots & 0 & \lambda
\end{array} \right) , $$
then by changing $ (e_j, e_{j+1}, \cdots , e_{j+p}) $ into $ (e_j, \varepsilon
e_{j+1} ,  \cdots , \varepsilon^{p} e_{j+p}) $  with $ \varepsilon >  0 $, the
above block is changed into the same  one with $ 1 $ replaced by $ \varepsilon
$. Proceeding similarly  for all the blocks, we obtain  the existence of basis
in which $ A $ is the sum of a diagonal matrix of norm $ e^{\Gamma_A} $ and of
a nilpotent matrix  of norm $ {\mathcal O} (\varepsilon) $.  This leads to the
statement when the spectrum is real. For non real eigenvalues $ \lambda = \rho
e^{i \theta}$, using Jordan normal form over $ \Ca^{2d} $, we have to consider
blocks of the form
$$ \left( \begin{array}{cc} J(\lambda) & 0 \\ 0 & J (\bar{\lambda})
\end{array} \right) .  $$
It is then standard  that there exists a basis of {\it  real} vectors in which
the endomorphism represented by the above block has a matrix of the form $ N +
\rho  R (\theta)  $ where  $ N  $ is  nilpotent and  $ R  (\theta) $  is block
diagonal matrix of rotations (of dimension $  2 $) of angle $ \theta $.  Then,
by changing this basis as in the case of a real spectrum, we can assume that $
N $ is small and we obtain $ \reff{nilpotent} $ in the general case.

\section{Equirepartition of time-evolved localized states}
\setcounter{equation}{0}
\subsection{The example of (generalized) coherent states}
In  this subsection,  we shall  prove  that the  generalized coherent  states,
defined below,  when evolved over  sufficiently long times,  equidistribute on
the torus.

To define the states in question, we proceed as follows. Let
\begin{eqnarray}
 \varphi_{\hbar}(q)  =  h^{-\mu  /  2}  \varphi  \left(  \frac{q}{\hbar^{\mu}}
 \right) \label{scaling}
\end{eqnarray}
with $ \varphi \in {\mathcal S}(\Ra^{d}) $, $ \int |\varphi|^2 = 1 $ and $ \mu
\in (0,1) $.
 Then we set
\begin{eqnarray}
 \varphi_{\hbar}^a = U_{\hbar}(a) \varphi_{\hbar} \label{notationetats}
\end{eqnarray}
  which  defines  a   family  of  states  in  $  L^2   (\Ra^d)  $  indexed  by
  $a\in\Ra^{2d}$.  These  are  commonly  referred to  as  (generalized)  coherent
  states. The corresponding
states on the torus, {\it i.e.} belonging to $ {\mathcal H}_{\hbar}(\kappa) $, are
defined by
\begin{eqnarray}
 \varphi_{\hbar,\kappa}^{a}  := S_{\hbar}(\kappa)  \varphi^a_{\hbar}  = \left(
\sum_{n_p  \in \Za^{d}} e^{-  i \kappa_q  \cdot n_p}  U_{\hbar}(0,n_p) \right)
\left(  \sum_{n_q  \in \Za^d}  e^{i  \kappa_p  \cdot  n_q} U  (n_q,0)  \right)
\varphi_{\hbar}^a
\label{projection}
\end{eqnarray}
which converges  in $ {\mathcal S}^{\prime}(\Ra^{d}) $  (see \cite{BoDB}). The
main property of these states that we shall use is
\begin{eqnarray}
\left\langle \varphi^a_{\hbar,\kappa} ,  O \! p^W (f)
\varphi^b_{\hbar,\kappa} \right\rangle_{  {\mathcal
H}_{\hbar}(\kappa)   }  =  \sum_{n  \in  \Za^{2d}} (-1)^{Nn_q
\cdot n_p}  e^{i \omega  (\kappa,n)} e^{i  \omega (n,b)/  2 \hbar}
\left\langle \varphi^a_{\hbar} , O \! p^W (f)
\varphi^{b-n}_{\hbar} \right\rangle_{L^{2}(\Ra^{d})}
 \label{Laformule}
\end{eqnarray}
which is proven in \cite{DBBo0}. The best known example of such
functions are obtained by choosing $ \mu  = 1  / 2 $  and $
\varphi (q)  = \eta  (q) := \pi^{-d/4} e^{-q^2 / 2} $. With this
choice one obtains the standard coherent states.

If $ \tilde{\varphi} $ is another Schwartz function
and $ \tilde{\varphi}_{\hbar}$ is defined similarly to $ \reff{scaling} $, the
Wigner  function  $  W_{\hbar}  (x)   $  associated  to  $  \varphi_{\hbar}  ,
\tilde{\varphi}_{\hbar} $ is defined by
$$  \left\langle  \varphi_{\hbar}  ,  O  \!  p^W  (f)  \tilde{\varphi}_{\hbar}
\right\rangle_{L^2(\Ra^d)} = \int_{\Ra^{2d}} f (x) W_{\hbar}(x) \ \mbox{d}x
$$ for all  $ f \in {\mathcal B}(\Ra^{2d}) $. For  general $ \varphi_{\hbar} ,
\tilde{\varphi}_{\hbar}  $  in  $  L^{2}(\Ra^{d})   $,  $  W_{\hbar}  $  is  a
distribution, but  for Schwartz  functions it is  a Schwartz function  as well
given by
$$ W_{\hbar} (x) =  (2 \pi \hbar)^{-d} \int e^{- i \tilde{q}  \cdot p / \hbar}
\overline{\varphi_{\hbar}  (q  -  \tilde{q} /  2)}  \tilde{\varphi}_{\hbar}(q+
\tilde{q}/2) \ \mbox{d}  \tilde{q} , \qquad x  = (q,p) . $$ With  the simple $
\hbar $  dependence considered in $ \reff{scaling}  $, it is easy  to see that
the Wigner  function $  W^{(a,b)}_{\hbar}(x) $ associated  to $  U_{\hbar} (a)
\varphi_{\hbar}  $ and  $ U_{\hbar}  (b) \tilde{\varphi}_{\hbar}  $  takes the
following form for any $ a,b \in \Ra^{2d} $
\begin{eqnarray}
 W_{\hbar}^{(a,b)} (x) =  e^{ - i \omega  (a,b)/ 2 \hbar + i  \omega (x,b-a) /
 \hbar} \hbar^{-d}  W_1 \left(  \Sigma^{\mu}_{\hbar} \left( x  - \frac{a+b}{2}
 \right) \right)\label{Wigneroff}
\end{eqnarray}
where $ \Sigma^{\mu}_{\hbar} $ is the linear  map on $ \Ra^{2d} $ defined by $
\Sigma_{\hbar}^{\mu} (q,p)  = (q  / \hbar^{\mu}  , p /  \hbar^{1-\mu} )  $ and
$W_1$ the Wigner function of $\varphi,\tilde{\varphi}$. Note that, since $ W_1
\in   L^1    (\Ra^{2d})   $,   $    \reff{Wigneroff}   $   implies    that   $
||W_{\hbar}^{(a,b)}||_{L^1}  = ||W_1  ||_{L^1}  $ is  independent  of $  \hbar
$. Note also that  when $ \varphi (q) = \tilde{\varphi} (q)  = \eta (q) $, one
easily checks that
\begin{eqnarray}
W_1 (x) = \pi^{-d} e^{-x^2} \label{gaussienne}
\end{eqnarray}
which makes  $ \reff{Wigneroff}  $ completely explicit  in this case.

Our main result is Theorem \ref{cohstateevol2}. As explained in the
introduction, its proof  goes in
two steps.  First we use  the Egorov theorem  to establish that on  a suitable
time scale $ \left\langle  U_{\epsilon}^t  \varphi_{\hbar,\kappa}^a  ,  O  \!  p^W  (f)
U_{\epsilon}^t        \varphi_{\hbar,\kappa}^a        \right\rangle_{{\mathcal
H}_{\hbar}(\kappa)} $   is equivalent
to $\left\langle  \varphi_{\hbar}^a  , O  \!  p^W (f  \circ
\Phi_{\epsilon}^t)  \varphi_{\hbar}^a \right\rangle_{L^2(\Ra^{d})} $
(Proposition \ref{CLASSIQUE}).
 Then we  use an
estimate on the classical evolution (exponential mixing) to control this last term.

As a warm up for the first step, we show for a particularly simple
class of states how  the  Egorov  expansion  $ \reff{expansion} $
can be reduced to the first term.

\begin{prop} \label{inutile} Let $  \Psi_{\hbar}  \in {\mathcal H}$ be a family such that there exists
$ C $ satisfying
\begin{eqnarray}
\left|  \left\langle \Psi_{\hbar} ,  O \!  p^W (f)  \Psi_{\hbar} \right\rangle
\right| \leq C || f ||_{\infty} , \qquad 0 < \hbar \leq 1 \label{mesure}
\end{eqnarray}
for all  $ f $ in ${\mathcal  B}({\mathcal M})$ having a  bounded and analytic
 continuation to some $ \Omega_{\delta} $. Then
$$  \left\langle U_{\epsilon}^t  \Psi_{\hbar} ,  O \!  p^W  (f) U_{\epsilon}^t
\Psi_{\hbar} \right\rangle_{{\mathcal  H}} - \left\langle \Psi_{\hbar}  , O \!
p^W  (f  \circ  \Phi_{\epsilon}^t) \Psi_{\hbar}  \right\rangle_{{\mathcal  H}}
\rightarrow 0  , \qquad \hbar  \rightarrow 0  $$ provided $  0 \leq t  \leq (2
-\nu) | \ln \hbar | / 3 \Gamma_{\epsilon} $ for some $ \nu \in (0,2) $.
\end{prop}

\noindent {\it Proof.} Using Theorem $ \refe{main} $, we only have to show that for all
$ 1 \leq j < J $ we have
$$  \hbar^j \left\langle  \Psi_{\hbar}  , O  \!  p^W (  {\mathcal  L}_j^t f  )
\Psi_{\hbar}  \right\rangle_{{\mathcal  H}}   \rightarrow  0  ,  \qquad  \hbar
\rightarrow 0  $$ in the specified  range of times. This  readily follows from
the fact that
$$ \hbar^j || {\mathcal L}_j^t  f ||_{\infty} \leq C_j \hbar^{j} |\ln \hbar|^j
e^{  - j  \left(  1 -  \frac{\nu}{2}  \right) \ln  \hbar }  $$  by
estimate  $ \reff{suiteCalderon}$, where one should recall that
${\mathcal L}_j^t={\mathcal K}^t_{j, J}$ if $j<J$. \finpreuve

\medskip

The condition  $ \reff{mesure} $ is for instance satisfied by
coherent states, in both cases $ {\mathcal M} = \Ra^{2d}$ and $
\Ta^{2d} $. This readily follows from the $ \hbar $ independence
of $ ||W^{(a,a)}_{\hbar} ||_{L^1}$ if $ {\mathcal M} = \Ra^{2d} $.
In case of the torus, it is a simple exercise using the Poisson
summation formula. Note also that, if $ f $ is periodic (in
particular if ${\mathcal M} = \Ta^{2d} $), we can get rid of the
analyticity of $  f $ since it is the uniform limit of a sequence
of trigonometric polynomials.

Nevertheless, regarding coherent states on the torus, the above
result is not precise enough for our purpose since the term $
\left\langle \varphi_{\hbar,\kappa}^a , O  \! p^W (f \circ
\Phi_{\epsilon}^t ) \varphi_{\hbar,\kappa}^a \right\rangle $ is
not very explicit. This is why we give the next proposition whose
proof will also be used in the proof of Theorem $
\refe{eigenfunctionsclassique} $.
\begin{prop} \label{CLASSIQUE}
  Fix $ a \in \Ra^{2d}  $ and assume that $ 0 < \mu < 1  $.  Then, for all $ f
   \in C^{\infty} ({\mathbb T}^{2d})$,
we have
$$  \left\langle  U_{\epsilon}^t  \varphi_{\hbar,\kappa}^a  ,  O  \!  p^W  (f)
U_{\epsilon}^t        \varphi_{\hbar,\kappa}^a        \right\rangle_{{\mathcal
H}_{\hbar}(\kappa)}  - \left\langle  \varphi_{\hbar}^a  , O  \!  p^W (f  \circ
\Phi_{\epsilon}^t)  \varphi_{\hbar}^a \right\rangle_{L^2(\Ra^{d})} \rightarrow
0 ,  \qquad \hbar \rightarrow 0 $$  provided $ 0 \leq  t \leq (2 -  \nu) | \ln
\hbar | / 3 \Gamma_{\epsilon} $ for some $ \nu \in (0,2) $.
\end{prop}

\noindent {\it Proof.} Let us first note that, by truncating the
Fourier series  of $  f $, there  exists a sequence  $ f_M  $ of $  \Za^{2d} $
periodic  analytic functions such  that $  f_M \rightarrow  f $  in ${\mathcal
B}(\Ta^{2d})$. Since  $ ||  O \!  p^W (f) -  O \!  p^W (f_M)  ||_{{\mathcal H}
\rightarrow {\mathcal H}} \rightarrow 0 $ and
$$  \left\langle \varphi_{\hbar}^a  ,  O \!  p^W  (f \circ  \Phi_{\epsilon}^t)
\varphi_{\hbar}^a       \right\rangle_{L^2(\Ra^{d})}       -      \left\langle
\varphi_{\hbar}^a ,  O \! p^W (f_M  \circ \Phi_{\epsilon}^t) \varphi_{\hbar}^a
\right\rangle_{L^2(\Ra^{d})} \rightarrow 0 , \qquad M \rightarrow + \infty
$$  uniformly with  respect to  $t  \in \Ra$  and $  \hbar  \in (0,1]  $ by  $
\reff{Wigneroff} $, we are  left with the case where $ f  $ is analytic. Then,
by Theorem $ \refe{main} $, we only have to study the difference
$$  \sum_{j <  J} \hbar^j  \left\langle  \varphi_{\hbar,\kappa}^a ,  O \!  p^W
 ({\mathcal   L}_j^t   f)  \varphi_{\hbar,\kappa}^a   \right\rangle_{{\mathcal
 H}_{\hbar}(\kappa) }  - \left\langle  \varphi_{\hbar}^a , O  \! p^W  (f \circ
 \Phi_{\e}^t) \varphi_{\hbar}^a \right\rangle_{L^{2}(\Ra^{d})  } , $$ thus the
 result  will  follow from  $  \reff{Laformule}  $ if  we  show  that, in  the
 specified range of times,
\begin{eqnarray}
\sum_{n  \ne 0}  \left|  \left\langle  \varphi_{\hbar}^a ,  O  \!p^W (f  \circ
\Phi^t_{\e})   \varphi^{a-n}_{\hbar}   \right\rangle_{L^2(\Ra^d)}  \right|   &
\rightarrow & 0 ,
\label{restel1}
 \\ \hbar^j  \sum_{n \in \Za^{2d}}  \left| \left\langle \varphi_{\hbar}^a  , O
\!p^W  ({\mathcal L}_j^t  f)  \varphi^{a-n}_{\hbar} \right\rangle_{L^2(\Ra^d)}
\right| & \rightarrow & 0 \qquad j \geq 1 . \label{sousreste}
\end{eqnarray}
We first  note that the term corresponding  to $ n=0$ in  $ \reff{sousreste} $
has been studied in the proof of  the previous proposition, and its limit is $
0  $.  We  may  therefore  assume  that  $n \ne  0$  in  both  sums.  Using  $
\reff{Wigneroff} $,  integrations by parts with  $ \hbar^2 \Delta_x  / |n|^2 $
show that, for all $j \geq 0$ and all $ M > 0 $
$$\left| \hbar^j \left\langle \varphi^{a}_{\hbar} , O \!p^w ({\mathcal L}_j^t f) \varphi^{a-
n }_{\hbar}  \right\rangle_{L^2(\Ra^d)} \right| \leq  C_j |n|^{- 2
M } \sum_{|\gamma| \leq 2  M} \hbar^{ j +  2 M -
\overline{m}(2M-|\gamma|)} || \partial^{\gamma} {\mathcal L}_j^t f
||_{\infty},
  $$
where $\overline m = \max(\mu, 1-\mu)$.
We get the result by the simple observation that
$$ \hbar^{ j + 2 M - \overline{m}(2M-|\gamma|)} || \partial^{\gamma} {\mathcal
L}_j^t f ||_{\infty}  \leq C_f \hbar^{2 \nu j  + C M } \rightarrow  0 , \qquad
\hbar \rightarrow 0
$$ for $ | \gamma | \leq 2 M$, with $ C = 2 (1+\nu)/3 $ if $ \overline{m} < (2
- \nu) / 3 $  and $ C = 2 (1 - \overline{m}) $  otherwise. This follows from $
\reff{suiteCalderon}  $  by  distinguishing  both cases  $  \overline{m}  \geq
(2-\nu)/3 $ and $ \overline{m} < (2-\nu)/3 $.
\finpreuve

\medskip

This proposition, combined  with $ \reff{Wigneroff} $ allows  us to reduce the
study of the matrix elements of evolved coherent states to a problem in classical
dynamics.
By this, we mean that the main result of this section, Theorem
\ref{cohstateevol2}, is a direct consequence   of  Proposition   $
\refe{CLASSIQUE} $ and of the mixing estimates given in the Appendix A.

Note that from now on, we shall be working with $d=1$. As explained in the
introduction, the reason for this is
that, whereas the mixing rate is controlled by the smallest Lyapounov exponent
of $A$, the error in the Egorov theorem is controlled by its largest Lyapounov exponent.

As a warm-up, and in order to bring out the main strategy, we first prove a
simplified version of the result:
\begin{theo} \label{Qmixing}  Assume that $ \Gamma_A > 0 $. Let $ a $ in $ \Ra^{2}
$, $ f \in {\mathcal B}(\Ta^2) $ and
$ 1/3 <  \mu < 2/3 $.  Then, for all  $ \nu > 0 $ there  exists $ \epsilon_0 $
 small enough (independent of $f$) such that for $ | \epsilon | < \epsilon_0 $
 we have
$$  \left\langle  U_{\epsilon}^t  \varphi^a_{\hbar,\kappa}  ,  O  \!  p^W  (f)
U_{\epsilon}^t        \varphi^a_{\hbar,\kappa}        \right\rangle_{{\mathcal
H}_{\hbar}(\kappa)} \rightarrow \int_{\Ta^2} f  (x) \ \emph{d}x , \qquad \hbar
\rightarrow 0 ,
 $$ provided that
\begin{eqnarray}
 \frac{  \overline{m} + \nu}{  \Gamma_{\epsilon}} |  \ln \hbar  | \leq  t \leq
\frac{2 - \nu}{3  \Gamma_{\epsilon}} |\ln \hbar| , \qquad  \overline{m} = \max
(\mu , 1 - \mu) . \label{souscondition}
\end{eqnarray}
\end{theo}

\noindent {\it Proof.} We first remark that, by choosing $ 0 < \Gamma < \Gamma_A $
close enough to $ \Gamma_A $ and $ \epsilon $ small enough we have
\begin{eqnarray}
 1 > \frac{\Gamma}{\Gamma_{\epsilon}} > \frac{1 + \nu / 2}{1 + \nu} .
\label{conditionouverte}
\end{eqnarray}
Combined with $ \reff{souscondition} $, this  estimate implies that $ t / |\ln
\hbar| > ( \overline{m} + \nu/2) / \Gamma $ and thus
\begin{eqnarray}
 e^{-t \Gamma} \leq \hbar^{\overline{m}+\frac{\nu}{2}} . \label{competition}
\end{eqnarray}
 By Proposition $ \refe{CLASSIQUE} $ and  $ \reff{Wigneroff} $ we only have to
study the limit of
\begin{eqnarray}
 \int_{\Ra^{2}} (f \circ \Phi_{\e}^t) (x) W^{(a,a)}_{\hbar}(x) \ \mbox{d}x
\label{Wignersurtore}
\end{eqnarray}
 for  which $  \int W^{(a,a)}_{\hbar}(x)  \mbox{d}x  = 1$.  Choosing a  smooth
cutoff function $ \chi $ so that $  \chi = 1 $ near $0$ and which is supported
close to $  0 $, then setting $  g_{\hbar}(x):= W^{(a,a)}_{\hbar}(x) \chi(x-a)
$,   we  have  $   ||  W^{(a,a)}_{\hbar}   -  g_{\hbar}||_{L^1}   =  {\mathcal
O}(h^{\infty}) $ thus
$$ \int_{\Ra^{2}}  (f \circ \Phi_{\e}^t) (x) W^{(a,a)}_{\hbar}(x)  \ \mbox{d}x -
\int_{\Ra^{2}} (f \circ \Phi_{\e}^t)  (x) g_{\hbar}(x) \ \mbox{d}x \rightarrow 0
, \qquad \hbar \downarrow 0
  $$ uniformly with respect to $ t  \in \Ra $. The last integral can obviously
be interpreted as an integral over $  \Ta^2 $ since $ g_{\hbar} $ is supported
close to $  a $ and consequently  we can use Corollary $  \refe{Mixing} $. The
result  now simply  follows  from  the fact  that  $e^{-t \Gamma}  ||g_{\hbar}
||_{W^{1,1}} = {\mathcal O}(h^{ -  \overline{m} }) e^{-t \Gamma} \rightarrow 0
$ by $ \reff{competition} $. \finpreuve

\bigskip

The   above  proof   is  a   rather  direct   application  of   Proposition  $
\refe{CLASSIQUE} $  and Corollary $ \refe{Mixing}  $ but it fails if $
\overline{m} \geq  2/3 $ ({\it  i.e.} $\mu \notin  (1/3,2/3)$) since $  e^{- t
\Gamma}  h^{- \overline{m}}  > 1  $, for,  in that  case, $  e^{- t  \Gamma} >
\hbar^{2/3}  $.  The problem stems from the lower bound in
(\ref{souscondition}), which arises because $\nabla W_\hbar^{(a,a)}$ behaves
like $\hbar^{-\overline m}$.  One
expects on intuitive grounds that it should be possible to replace $\overline
m$ by
$\underline m=\min(\mu, 1-\mu)$ which is of course less than $1/2$ which is
less than $2/3$. We shall  prove this is true, but  for that purpose we will
need to exploit some  more detailed knowledge about the Anosov diffeomorphisms we study.
The trick consists in applying a well known idea in the theory of Anosov
systems: it is possible to replace (\ref{Wignersurtore}) by an expression obtained by performing
an integral along the stable foliation. Since the evolution stretches the
function $W_\hbar^{(a,a)}$ along the unstable manifold, this corresponds to smoothening out
the fastest oscillations in $W_\hbar^{(a,a)}$, replacing the latter by a function that has a
derivative controlled by $\hbar^{-\underline m}$.
Let us start the proof. By  Proposition $ \refe{CLASSIQUE} $, we have to study
$ \reff{Wignersurtore}  $ where $ W^{(a,a)}_{\hbar}  $ can be  replaced, as in
the proof of Theorem $ \refe{Qmixing} $,  by $ g_{\hbar} $ which we can assume
to be supported as close to $ a $ as we want. This will allow us to use the following
result.
\begin{theo}\cite{BKL1,BrSt} \label{coordonnees} For all $ \Gamma < \Gamma_A $, there exists $ \epsilon_0 $
small  enough such  that for  all $  |\epsilon| <  \epsilon_0 $
the following holds:  there  exist $  \sigma_{\epsilon} > 0 $ and
a $ C^{1+ \sigma_{\epsilon}} $ diffeomorphism  $  x  \mapsto
F_{\epsilon}(x) = (s(x),u(x)) $,  from a neighborhood of $ a \in
\Ta^2$ to a neighborhood of  $ 0 \in \Ra^2 $ such that $
F_{\epsilon}(a) = 0 $ and
\begin{eqnarray}
\left| \partial_s  \left( f \circ \Phi_{\e}^t \circ  F_{\e}^{-1}
\right) (u,s) \right| \leq C_f e^{- \Gamma t} \label{Taylorexp}
\end{eqnarray}
for all $ t \geq  0 $, all $ (u,s) $ in the neighborhood of $  0 $ and all $ f
\in  C^1  (\Ta^2,\Ra) $.  Here  $  C^{1+\sigma}  $ denotes  the  corresponding
H\"older class.
\end{theo}
Using this result, we can perform the following change of variables
\begin{eqnarray}
 \int_{\Ra^2} \left( f \circ \Phi^t_{\e} \right)(x) g_{\hbar}(x) \ \mbox{d}x =
\int \!  \!  \int \left( f  \circ \Phi^t_{\e} \circ  F_{\e}^{-1} \right) (u,s)
\left(  g_{\hbar}  \circ  F_{\epsilon}^{-1} \right)(u,s)  J_{\epsilon}(u,s)  \
\mbox{d}u \mbox{d}s \label{variables}
\end{eqnarray}
where $ J_{\epsilon}  \in C^{\sigma_{\epsilon}} $. On the right
hand side of this
  equation,
we eventually want  to use Corollary $ \refe{Mixing} $, but the $
C^{\sigma_{\epsilon}}  $ regularity of $J_\epsilon(u,s)$
 is not sufficient for that purpose. Fortunately, the term $  J_{\epsilon} $
is essentially irrelevant in view
of the following result.
\begin{lemm} \label{regulJacobian} i) $(g_{\hbar} \circ F_{\e}^{-1})_{0<\hbar
  \leq 1}$ is a bounded family in  $ L^1(\Ra^2) $.
 \newline  ii) For all  $ \theta \in (0,1)  $ there
exists a family $ J_{\e}^{\hbar} $ such  that, if $ || .  ||_{\infty} $ is the
$ \sup $ norm over a fixed small neighborhood of $ 0 $,
$$  || J_{\e}^{\hbar}  - J_{\e}  ||_{\infty}  \leq C  \hbar^{\sigma_{\epsilon} \theta}  ,
\qquad || \nabla J_{\e}^\hbar ||_{\infty} \leq C \hbar^{-\theta} .
$$
\end{lemm}

\medskip

\noindent {\it Proof.} {\it i)} follows from $ \reff{Wigneroff} $
and  {\it ii)}  from a  standard convolution  argument by  a $  C^{\infty}_0 $
function $ \chi_{\hbar} (u,s) = \hbar^{-2  \theta} \chi ( u / \hbar^{\theta} ,
s / \hbar^{\theta})$. \finpreuve

\bigskip

Using  this  lemma  and  $  \reff{Taylorexp}  $, the  right  hand  side  of  $
\reff{variables} $ takes the form
$$  \int  \left(  f  \circ   \Phi^t_{\e}  \circ  F_{\e}^{-1}  \right)  (u,  0)
k_{\hbar}(u) \  \mbox{d}u + {\mathcal O_1}  (e^{-\Gamma t} )  +
{\mathcal O}_2 (\hbar^{\sigma_{\epsilon} \theta})
$$ where $ {\mathcal O}_1 $ is uniform  with respect to $ \hbar \in (0,1] $, $
{\mathcal O}_2 $ is uniform with respect to $ t \geq 0 $ and where the $ C^1 $
function $ k_{\hbar}(u) $ is given by
$$ k_{\hbar}(u)  = \int \left( g_{\hbar}  \circ F_{\epsilon}^{-1} \right)(u,s)
J_{\epsilon}^{\hbar}(u,s) \ \mbox{d}s .  $$ Note that $ k_{\hbar}
$ is bounded in $ L^1 $ and that $  \int_{\Ra} k_{\hbar} (u) \
\mbox{d}u \rightarrow 1 $ as $ \hbar  \rightarrow 0 $. The key
remark is now that the derivative of this function is essentially
controlled by $\hbar^{-\underline m}$ rather than by
$\hbar^{-\overline m}$, as a rough estimate would show. That is
the content of the following proposition. Note that, in what
follows, $ q,p $ are the canonical  coordinates of $ \Ra^2 $. They
also define local  coordinates on  $ \Ta^2  $ close to any $  a $,
and  this makes  the following statement clear.
 \begin{prop} Assume that $ \mu \leq 1/2 $ ({\it i.e.} that
 $\max(\mu,1-\mu) =  1 -  \mu $).  Then, if the  support of  $ g_{\hbar}  $ is
 sufficiently close to $ a $ and if
\begin{eqnarray}
\partial_s \left( p \circ F_{\e}^{-1} \right) (0,0) \ne 0
\label{deriveenonnulle}
\end{eqnarray}
  then, there  exists $ \tilde{k}_{\hbar} \in  C^1 (\Ra) $ such  that $ \left|
  \left| k_{\hbar}  - \tilde{k}_{\hbar} \right| \right|_{L^1}  \rightarrow 0 $
  as $ \hbar \rightarrow 0 $ and
\begin{eqnarray}
 \left| \left| d  \tilde{k}_{\hbar} /du \right| \right|_{L^1} \leq  C \hbar^{- \mu }
. \label{powerinf}
\end{eqnarray}
 \end{prop}

 \medskip

\noindent {\it Proof.} The condition $ \reff{deriveenonnulle} $
shows that,  if $ \delta_1,\delta_2 $ are  small enough, $ s  \mapsto (p \circ
F_{\e}^{-1})(u,s) $ is a diffeomorphism from $ (-\delta_1,\delta_1) $ onto its
range  for each $  u \in  (-\delta_2,\delta_2) $.  Thus, if  the support  of $
g_{\hbar} $ is small enough, we can use $(p \circ F_{\e}^{-1})(u,s) $ as a new
variable in the integral defining $ k_{\hbar} $ so that it becomes
$$     k_{\hbar}     (u     )     =     \hbar^{-1}     \int     W_1     \left(
\frac{\tilde{q}(u,p)-q(a)}{\hbar^{\mu}} , \frac{p-p(a)}{\hbar^{1-\mu}} \right)
\widetilde{\chi}_{\epsilon}^{\hbar}(u,p) \ j_{\epsilon}(u,p) \mbox{d} p
$$ with  $ j_{\epsilon}(u,p) $  the $ C^{\sigma_{\epsilon}}  $ jacobian of the  change of
variable   and   $   \widetilde{\chi}_{\epsilon}^{\hbar}(u,p)   $   the   term
corresponding   to  $   \chi   (F_{\e}^{-1}(u,s)-a)  J_{\epsilon}^{\hbar}(u,s)
$. Changing again  the variable with $ \tilde{p}  = (p-p(a))/ \hbar^{1-\mu} $,
we would get the result if $ j_{\epsilon}  $ was $ C^1 $, by choosing $ \theta
= \mu $.  We can overcome the non  smoothness of $ j_{\epsilon} $  by the same
principle   as   for   Lemma   $   \refe{regulJacobian}   $:   we   choose   $
j_{\epsilon}^{\hbar} $  approaching $ j_{\epsilon} $ uniformly  on the support
of   $    \widetilde{\chi}_{\epsilon}^{\hbar}   $,   such    that   $   \nabla
j_{\epsilon}^{\hbar} = {\mathcal O} (\hbar^{-\mu})$ and then
$$     \tilde{k}_{\hbar}(u)      =     \hbar^{-1}     \int      W_1     \left(
\frac{\tilde{q}(u,p)-q(a)}{\hbar^{\mu}} , \frac{p-p(a)}{\hbar^{1-\mu}} \right)
\widetilde{\chi}_{\epsilon}^{\hbar}(u,p)  \ j_{\epsilon}^{\hbar}(u,p) \mbox{d}
p $$ has the expected properties. \finpreuve

\bigskip

\noindent {\bf Remark.} The condition $ \reff{deriveenonnulle} $
expresses the fact that, at the point $ a $, the submanifold $ \{ q = q (a) \}
$ is not aligned  with the {\it unstable manifold}. Of course,  if $ \mu > 1/2
$, the same result holds if $ \partial_s (q \circ F_{\e}^{-1} )(0,0) \ne 0 $.

\bigskip

We are now ready for the proof of the main  theorem of this subsection.

\begin{theo}\label{cohstateevol2} Assume that $ 0 < \mu \leq 1/3 $ (resp. $ 2/3 \leq \mu < 1
$)  and that  the unstable  manifold through  $ a  $ is  not aligned  with the
submanifold $  \{ q = q(a) \}  $ (resp. $\{ p  = p (a) \}  $). Assume moreover
that $  \Gamma_A > 0  $. Then, there  exists $ \epsilon_0  $ such that,  for $
|\epsilon| < \epsilon_0 $ and all $ f \in C^{\infty}(\Ta^2) $
$$  \left\langle  U_{\epsilon}^t  \varphi^a_{\hbar,\kappa}  ,  O  \!  p^W  (f)
U_{\epsilon}^t        \varphi^a_{\hbar,\kappa}        \right\rangle_{{\mathcal
H}_{\hbar}(\kappa)} \rightarrow \int_{\Ta^2} f  (x) \ \emph{d}x , \qquad \hbar
\rightarrow 0 ,
 $$ provided that
\begin{eqnarray}
 \frac{ \underline{m}  + \nu}{  \Gamma_{\epsilon}} | \ln  \hbar | \leq  t \leq
\frac{2 - \nu}{3 \Gamma_{\epsilon}} |\ln  \hbar| , \qquad \underline{m} = \min
(\mu , 1 - \mu) . \label{souscondition2}
\end{eqnarray}
\end{theo}

\noindent {\it Proof.} The above discussion shows that we only
have to prove that
\begin{eqnarray}
 \int   \left(  f  \circ   \Phi^t_{\e}  \circ   F_{\e}^{-1}  \right)   (u,  0)
\tilde{k}_{\hbar}(u)    \    \mbox{d}u    \rightarrow    \int_{\Ta^2}f(x)    \
\mbox{d}x. \label{reduction1d}
\end{eqnarray}
Pick a smooth  function $ \varrho (s) $  supported close to $ 0 $  such that $
\int \varrho (s)  \mbox{d}s =1 $. Then, using  Theorem $ \refe{coordonnees} $,
the left hand side of $ \reff{reduction1d} $ takes the form
$$ \int \! \! \int \left( f \circ \Phi^t_{\e} \circ F_{\e}^{-1} \right) (u, s)
\tilde{k}_{\hbar}(u)   \varrho(s)  \   \mbox{d}u  \mbox{d}s   +   {\mathcal  O
}(e^{-\Gamma_\epsilon t})
$$ with $ {\mathcal  O }(e^{-\Gamma_\epsilon t}) $ uniform with respect  to $ \hbar \in
(0,1] $. This last integral is nothing but
$$ \int_{\Ta^2} f \circ \Phi_{\epsilon}^t (x) \tilde{g}_{\hbar}(x) \ \mbox{d}x
$$ where  $ \tilde{g}_{\hbar}  \circ F_{\e}^{-1} (u,s)  = \tilde{k}_{\hbar}(u)
\varrho(s)/J_{\epsilon}(u,s) $.  Thus $ \tilde{g}_{\hbar}  $ is of
the  form $ \tilde{g}^{(1)}_{\hbar}   \tilde{g}^{(2)}  $   with $
\tilde{g}^{(2)}   \in C^{\sigma_{\epsilon}} $ independent of $
\hbar $ and $ ||\tilde{g}^{(1)}_{\hbar}||_{L^1}+
\hbar^{\underline{m}} || \nabla \tilde{g}^{(1)}_{\hbar} ||_{L^1} =
{\mathcal O}(1) $. Note also that $  \int_{\Ta^2}
\tilde{g}_{\hbar} \rightarrow 1 $ as $ \hbar \rightarrow 0 $.
Using  Lemma $ \refe{regulJacobian} $ again to approach $
\tilde{g}^{(2)}   $  by  $   C^1  $ functions, we may  assume that
$ \tilde{g}_{\hbar}  $  is   $ C^1  $ and satisfies  the same
bound  as  $ \tilde{g}^{(1)}_{\hbar} $. We can  now  repeat the
arguments  of Theorem  $ \refe{Qmixing} $ and the result follows.
\finpreuve

\subsection{Semiclassical behavior of eigenstates}
We now come to a more general result having applications in the description of
the eigenvectors  of  $ U_{\epsilon}  $.  Assume  that  $ \Psi_{\hbar,\kappa}  \in
{\mathcal H}_{\hbar}(\kappa) $ satisfies,  for all $ f \in C^{\infty}({\mathbb
T}^{2d}) $,
\begin{eqnarray}
 \left\langle   \Psi_{\hbar,\kappa},   O   \!  p^W   (f)   \Psi_{\hbar,\kappa}
\right\rangle_{ {\mathcal H}_{\hbar}(\kappa) } \rightarrow f (0), \qquad \hbar
\downarrow 0 .
\label{concentrationzero}
\end{eqnarray}
Rather vaguely, this condition says that $ \Psi_{\hbar,\kappa}$ is
concentrated at $ 0 $. This is confirmed by the following
\begin{lemm} \label{spectralcutoff} 
There exists a sequence of positive  numbers $ r_{\hbar} \rightarrow 0 $ and a
family of functions $ \chi_{\hbar}  \in C^{\infty} (\Ta^{2d}) $ supported in a
ball of radius $  r_{\hbar} $ centered at $ 0 $ (in  $\Ta^{2d}$) such that $ 0
\leq \chi_{\hbar} \leq 1 $ and
\begin{eqnarray}
 \left|  \left|  \Psi_{\hbar,\kappa}  -  (2  \pi  \hbar)^{-d}  \int_{\Ta^{2d}}
\chi_{\hbar}(a)    \left\langle   \eta^a_{\hbar,\kappa},   \Psi_{\hbar,\kappa}
\right\rangle_{   {\mathcal  H}_{\hbar}(\kappa)   }   \eta^a_{\hbar,\kappa}  \
\emph{d}a  \right|  \right|_{{\mathcal  H}_{\hbar}(\kappa)}  \rightarrow  0  ,
\qquad \hbar \downarrow 0 . \label{speed}
\end{eqnarray}
Conversely, if $ \reff{speed} $ holds and $
||\Psi_{\hbar,\kappa}||_{{\mathcal H}_{\hbar}(\kappa)} \rightarrow
1 $ then $ \reff{concentrationzero} $ holds for all $ f \in
C^{\infty}(\Ta^{2d}) $.
\end{lemm}

The proof  of this lemma is  given in Appendix  $ \refe{decompositionetats} $,
where we also recall basic results on the coherent states decomposition over $ L^2
(\Ra^d)   $   and   $   {\mathcal   H}_{\hbar}(\kappa)  $.   Recall   that   $
\eta^a_{\hbar,\kappa}   $    is   defined    by   $   \reff{scaling}    $,   $
\reff{notationetats} $ and $ \reff{projection} $ with $ \mu = 1/2 $ and $ \eta
(q) = \pi^{-d/4} e^{-q^2/2} $.

The right hand side in (\ref{concentrationzero}) 
could of course be replaced  by $ f (a_0) $ for some $ a_0 \in
\Ta^{2d} $ or more generally by $ \sum_{0 \leq j \leq J} \alpha_j f (a_J) $
for finitely many points $ a_0 , \dots , a_J  $. 
Correspondingly, one can then define the concentration on a finite collection of points
in a $ r_{\hbar} $ neighborhood of those points.

 To simplify the notation, we  set  $  \lambda_{\hbar}  (a)  =
\chi_{\hbar}(a)   \left\langle  \eta^a_{\hbar,\kappa}   ,  \Psi_{\hbar,\kappa}
\right\rangle_{ {\mathcal  H}_{\hbar}(\kappa) } $.  The above  lemma proves
that
$$ \psi_{\hbar,\kappa}  := (2 \pi  \hbar)^{-d} \int_{\Ta^{2d}} \lambda_{\hbar}
(a) \eta^a_{\hbar,\kappa} \  \mbox{d}a $$ satisfies $ \reff{concentrationzero}
$ as well and that
$$   \left\langle   \Psi_{\hbar,\kappa},   U_{\epsilon}^{-t}   O   \!   p^W(f)
U_{\epsilon}^t       \Psi_{\hbar,\kappa}       \right\rangle_{       {\mathcal
H}_{\hbar}(\kappa) }  - \left\langle \psi_{\hbar,\kappa},  U_{\epsilon}^{-t} O
\!   p^W(f)  U_{\epsilon}^t   \psi_{\hbar,\kappa}   \right\rangle_{  {\mathcal
H}_{\hbar}(\kappa) } \rightarrow 0, \qquad \hbar \downarrow 0
$$ uniformly with respect to $ t \geq 0 $. This is the first step of the proof
of the next  theorem, in which the notations  $ \langle . , .
\rangle $ and $ ||.|| $ stand  for $ \langle . , .
\rangle_{{\mathcal H}_{\hbar}(\kappa)} $ and $ ||.||_{{\mathcal
H}_{\hbar}(\kappa)}  $ respectively.
\begin{theo} \label{eigenfunctionsclassique} Assume that $
|| \Psi_{\hbar,\kappa} || \rightarrow 1 $ and that $ \reff{speed}
$ holds for some sequence $ r_{\hbar} $ such that
$$ r_{\hbar} \leq  \hbar^{1/2 - \sigma} , $$  with $ \sigma > 0 $.  Then, as $
\hbar \rightarrow 0 $,
$$   \left\langle   \Psi_{\hbar,\kappa},   U_{\epsilon}^{-t}   O   \!   p^W(f)
U_{\epsilon}^t  \Psi_{\hbar,\kappa}   \right\rangle  -  (2   \pi  \hbar)^{-2d}
\int_{\Ta^{2d}}    \int_{\Ta^{2d}}    \overline{    \lambda_{\hbar}   (a)    }
\lambda_{\hbar}  (b) \left\langle  \eta^a_{\hbar,\kappa}  , O  \! p^W(f  \circ
\Phi_{\epsilon}^t)  \eta^b_{\hbar,\kappa} \right\rangle \  \emph{d}a \emph{d}b
\rightarrow 0
$$ provided
\begin{eqnarray}
  0  \leq \Gamma_{\epsilon}  t \leq  \left(  \frac{1}{2} +  \tau \right)  |\ln
\hbar|, \qquad \frac{1}{2} - 3 \tau -  4 d \sigma > 0 \qquad \mbox{and} \qquad
\tau < \frac{1}{6} . \label{firstcondition}
\end{eqnarray}
If moreover $ d=1 $, $ \Gamma_A > 0 $  and $ \tau - 5 \sigma > 0 $, then there
exists $ t_{\hbar} \rightarrow \infty $ and $ \epsilon (\sigma,\tau) > 0$ such
that for all $ |\epsilon| \leq \epsilon (\sigma,\tau) $
\begin{eqnarray}
\left\langle  \Psi_{\hbar,\kappa},   U_{\epsilon}^{-t_{\hbar}}  O  \!   p^W(f)
U_{\epsilon}^{t_{\hbar}}    \Psi_{\hbar,\kappa}    \right\rangle   \rightarrow
\int_{\Ta^2} f (x) \ \emph{d}x, \qquad \hbar \downarrow 0.
\end{eqnarray}
\end{theo}

This theorem generalizes a result of \cite{DBBo1bis}, Section 5,
where only the case $\epsilon=0$ is treated. The proof is then
much simpler, since there is then no error term in the Egorov
theorem. The theorem  says that, if a sequence of states
concentrates sufficiently fast on a point $a$ in $\Ta^2$, then the
time evolved states equidistribute on the torus on some
logarithmic time scale. Before  proving this theorem,  we show how
it leads to a result on the semiclassical behaviour of the
eigenvectors  of $U_{\epsilon} $.

\begin{coro}\label{cor:eigenfunctionsclassique} Assume that $ d = 1 $ and that $ \Gamma_A > 0 $.
For any $  0< \sigma < 1 / 38  $, there exists $ \epsilon (\sigma)  > 0 $ such
that   for  all   $   |\epsilon|  <   \epsilon   (\sigma)  $,   no  family   $
\Psi_{\hbar,\kappa}  $  of  eigenvectors  of  $  U_{\epsilon}  $  can  satisfy
simultaneously $ \reff{concentrationzero} $ for all $ f $ and $ \reff{speed} $
with $ r_{\hbar} \leq \hbar^{1/2 - \sigma} $.
\end{coro}

We note in passing that a similar result (with a worse value of $\sigma$) holds
for $ d > 1 $ provided we impose a pinching condition on the Lyapounov exponents of $ A $
as mentioned in the introduction.

Roughly speaking,  this corollary  shows that, if a  family of eigenvectors  of $
U_{\epsilon} $  concentrates on a single point in
phase space in the semiclassical limit, then it must do so slowly. In other words, no such sequence  can
`live' in a ball of too small a radius $r_\hbar$.
In view of the comment after Lemma $ \reff{spectralcutoff}  $, 
it is clear that this result 
holds also for a pure point measure supported on a finite number of periodic orbits.          
 Given Theorem
\ref{eigenfunctionsclassique}, the
proof is very simple and
identical to the case $\epsilon=0$ treated in \cite{DBBo1bis}, Section 5. 
We repeat it for completeness.

\medskip

\noindent {\it Proof.}  For any $ 0 < \sigma < 1 / 3 8 $, one can
find $ \tau  > 5 \sigma $ satisfying  $ \reff{firstcondition} $.  Furthermore,
since   $   \Psi_{\hbar,\kappa}$   is   an   eigenfunction,   $   \left\langle
\Psi_{\hbar,\kappa},   U_{\epsilon}^{-t}    O   \!   p^W(f)   U_{\epsilon}^{t}
\Psi_{\hbar,\kappa}  \right\rangle =  \left\langle  \Psi_{\hbar,\kappa}, O  \!
p^W(f) \Psi_{\hbar,\kappa} \right\rangle $ for all $ t $, thus by choosing $ t
= t_{\hbar} $ and letting $ \hbar \downarrow 0 $ we obtain
$$   f  (0)  =   \int_{\Ta^2}  f(x)   \  \mbox{d}x   $$  for   all  $   f  \in
C^{\infty}(\Ta^{2d}) $, which leads to a contradiction. \finpreuve

\bigskip

\noindent {\bf Proof of Theorem ${\bf
\refe{eigenfunctionsclassique}}$.} Here again, it is sufficient to assume that
$   f  $   is  analytic.   Using  Theorem   $  \refe{main}   $  and   Lemma  $
\refe{spectralcutoff}  $, it  is  clear  that, if  $  \tau <  1/6  $  and $  t
\Gamma_{\epsilon} \leq (1/2 + \tau) |\ln \hbar| $, we have
$$   \left\langle   \Psi_{\hbar,\kappa},   U_{\epsilon}^{-t}   O   \!   p^W(f)
U_{\epsilon}^t  \Psi_{\hbar,\kappa}  \right\rangle  -  \sum_{j  <  J}  \hbar^j
\left\langle   \psi_{\hbar,\kappa}   ,  O   \!   p^W   ({\mathcal  L}_j^t   f)
\psi_{\hbar,\kappa} \right\rangle \rightarrow 0, \qquad \hbar \downarrow 0 .
$$ The first part of the theorem will  thus be proven if we show that, for any
$ j \geq 2 $ (recall that ${\mathcal L}_j^t \equiv 0$ if $j$ is odd), we have
$$    (2   \pi    \hbar)^{-2d}    \hbar^j   \int_{\Ta^{2d}}    \int_{\Ta^{2d}}
\overline{\lambda_{\hbar}(a)}       \lambda_{\hbar}      (b)      \left\langle
\eta^a_{\hbar,\kappa}  , O \!   p^W({\mathcal L}_j^t  f) \eta^b_{\hbar,\kappa}
\right\rangle \ \mbox{d}a \mbox{d}b \rightarrow 0, \qquad \hbar \downarrow 0
$$ if  $ 1/2 - 3  \tau - 4 d  \sigma > 0 $.  Using $ \reff{Laformule}  $ and $
\reff{Wigneroff} $,  integrations by parts  similar to those of  proposition $
\refe{CLASSIQUE} $ show easily that, for all $ M > 0 $,
$$ \left\langle \eta^a_{\hbar,\kappa} , O  \! p^W ({\mathcal L}_j^t  f )
\eta^b_{\hbar,\kappa} \right\rangle = \sum_{|n| \leq C} (-1)^{N n_q \cdot n_p}
e^{i \omega (\kappa,n) + i \omega (n,b)/ 2 \hbar } \left\langle \eta^a_{\hbar}
, O \! p^W ({\mathcal L}_j^t  f ) \eta^{b-n}_{\hbar} \right\rangle_{L^2}
+ {\mathcal O}(\hbar^{M}) $$
 uniformly with  respect to $ a,b \in [0,1)^{2d} $
and $ \Gamma_{\epsilon} t \leq (1/2 + \tau)|\ln  \hbar| $, with $ \tau < 1 / 6
$.  The constant  $ C  $  involved in  the sum  is  such that  $ |b-n-a|  \geq
C^{-1}|n| $ for all  $a,b \in [0,1)^{2d}$ and $ |n| > C  $. On the other hand,
using $  \reff{Wigneroff} $ and $  \reff{suiteCalderon} $, one  sees that, for
any $ n \in \Za^{2d} $ and any $ j \geq 2 $,
$$ (2  \pi \hbar)^{-2d} \int_{\Ta^{2d}}  \int_{\Ta^{2d}}\left| \lambda_{\hbar}
 (a) \lambda_{\hbar}(b) \right| \hbar^j \left| \left\langle \eta^a_{\hbar} , O
 \! p^W({\mathcal  L}_j^t f)  \eta^{b-n}_{\hbar} \right\rangle \right| \ \mbox{d}a \mbox{d}b
  \leq C \hbar^{-2d} r_{\hbar}^{4d} \hbar^{1/2 - 3 \tau}
$$ since $  |\lambda_{\hbar} (a) | \leq ||  \Psi_{\hbar,\kappa}|| $ is bounded
and  $  \lambda_{\hbar}  $  is  supported  in a  set  of  volume  $  {\mathcal
O}(r_{\hbar}^{2d})$. The first part of the theorem follows.

We now  prove the second part.   Since $ \chi_{\hbar}  $ can be chosen  of the
 form   $   \chi_{\hbar}(a)   =    \sum_{n_1   \in   \Za^{2d}}   \chi   \left(
 \frac{a+n_1}{r_{\hbar}}      \right)       $      (see      the      Appendix
 $\refe{decompositionetats}$),  it  turns   out  that,  for  any  $   M  $,  $
 \chi_{\hbar}(a) \chi_{\hbar}(b) \left\langle \eta^a_{\hbar,\kappa} , O \! p^W
 (f  \circ  \Phi_{\epsilon}^t )  \eta^b_{\hbar,\kappa}  \right\rangle$ can  be
 written
\begin{eqnarray*}
 \sum_{|b-a-n| =  {\mathcal O}(r_{\hbar})} (-1)^{N n_q \cdot  n_p} e^{i \omega
(\kappa,n) + i \omega (n,b)/ 2  \hbar } \left\langle \eta^a_{\hbar} , O \! p^W
(f   \circ  \Phi_{\epsilon}^t   )  \eta^{b-n}_{\hbar}   \right\rangle_{L^2}  +
{\mathcal O}(\hbar^{M})
\end{eqnarray*}
  uniformly with respect to $ a,b \in [ 0,1)^{2d} $. Now, if $ d = 1 $ and $ 5
  \sigma < \tau $, using $  \reff{Wigneroff} $ and proceeding similarly to the
  proof  of Theorem  $ \refe{Qmixing}$,  we see  that for  $ \epsilon  $ small
  enough and $ \Gamma < \Gamma_A $ sufficiently close to $ \Gamma_A $
$$  \left\langle  \eta^a_{\hbar} ,  O  \!  p^W  (f \circ  \Phi_{\epsilon}^t  )
\eta^{b-n}_{\hbar}   \right\rangle_{L^2}  -   \left\langle   \eta^a_{\hbar}  ,
\eta^{b-n}_{\hbar}  \right\rangle_{L^2} \int_{\Ta^2} f  (x) \mbox{d}x  = e^{-t
\Gamma} {\mathcal O}( \hbar^{-1/2} + r_{\hbar}/\hbar )
$$ uniformly on  the set where $  |b-n-a| = {\mathcal O}( r_{\hbar})  $, $ a,b
\in [ 0 , 1 )^2 $. This shows that
$$ (2 \pi  \hbar)^{-2} \int_{\Ta^2} \int_{\Ta^2} \overline{\lambda_{\hbar}(a)}
\lambda_{\hbar}(b)  \left\langle \eta^a_{\hbar,\kappa}  ,  O \!  p^W (f  \circ
\Phi_{\epsilon}^t  ) \eta^b_{\hbar,\kappa} \right\rangle  \mbox{d}a\mbox{d}b -
\left| \left| \psi_{\hbar,\kappa} \right| \right|^2 \int_{\Ta^2}f(x) \mbox{d}x
= {\mathcal O}( e^{- t \Gamma} \hbar^{- 1/2 - 5 \sigma } )
$$ and the result follows. \finpreuve
\appendix

\section{A mixing theorem for perturbations of hyperbolic maps on $ \Ta^{2}$.}
Let  $ A  $ be  a $  2 \times  2 $  matrix with  integer entries  such  that $
|\mbox{tr} A  | > 2$  and $ \mbox{det}A  = 1$. For notational  convenience, we
assume that its eigenvalues are  positive and we note them $e^{\pm \Gamma_A}$,
with  $  \Gamma_A  >  0  $.   Let  $  \phi_{\e}  $  be  a  measure  preserving
diffeomorphsim on $ \Ta^2 $, depending on a parameter $ \e $, such that
$$  \phi_{\e}  \rightarrow \  \mbox{id}  \qquad  \mbox{in}  \ C^3(\Ta^2)  \  \
\mbox{as}  \   \e  \rightarrow  0  .    $$  We  define   the  associated  {\it
Ruelle-Perron-Frobenius} operator $ {\mathcal L}_{\e} $ as the map
$$  {\mathcal L}_{\e} g  := g  \circ T_{\e}^{-1},  \qquad T_{\e}:  = \phi_{\e}
\circ A .  $$ Using \cite{BKL1} (more precisely $(2.1.7)$, Example $2.2.6$ and
Theorem $3$) one has the following result.
\begin{theo}[\cite{BKL1}] For any $ \Gamma < \Gamma_A  $, one can find $ \epsilon_0 > 0 $ small enough
such that  the following  property holds: for  all $  |\e| \leq \e_0  $, there
exists a Banach space $ {\mathcal  B}_{\epsilon} $ of distributions of order $
1 $, containing $C^1(\Ta^2)$, with norm $ ||.||_{\e} $ such that
$$ ||g||_{\e} \leq C_{\e} ||g||_{W^{1,1}}, \qquad \forall \ f \in C^1(\Ta^2)
$$ (with $||g||_{W^{1,1}}  = \int_{\Ta^2} |g| + \int_{\Ta^2}  |\nabla g|$) and
such that $$ {\mathcal L}_{\e}  = \Pi_1 + {\mathcal R}_{\e} \qquad \mbox{with}
\qquad {\mathcal R}_{\e} \Pi_1 = \Pi_1  {\mathcal R}_{\e} = 0 $$ where $ \Pi_1
g  = \scal{g,1}1  $ and  $ {\mathcal  R}_{\e}  $ is  a bounded  operator on  $
{\mathcal B}_{\e} $ with spectral radius lower  than $ e^{- \Gamma } $. Here $
\scal{.,.} $ is the  pairing between distributions of order $ 1  $ and $ C^1 $
functions.
\end{theo}
As a direct consequence, we obtain
\begin{coro} \label{Mixing} For all $ \Gamma < \Gamma_A $, there exists $ \e_0
  $
such that, for all $ |\e| < \e_0
$, one can find $ C_{\e,\Gamma} $ satisfying
$$  \left| \int_{\Ta^{2}}  f \left(  T_{\e}^t (x)  \right) g  (x)  \emph{d}x -
\int_{\Ta^{2}}  f \int_{\Ta^2}  g \right|  \leq C_{\e,\Gamma}  e^{-  t \Gamma}
||f||_{C^1}||g||_{W^{1,1}} , \qquad  \mbox{for all} \ f,g \in  C^1(\Ta^2), \ t
\geq 0 . $$
\end{coro}

\section{Generalized coherent states decompositions} \label{decompositionetats}
\setcounter{equation}{0} In  this appendix, we briefly recall  some results on
coherent states decompositions as well  as some convenient tools for the proof
of Lemma $ \refe{spectralcutoff} $.

As it is for instance proven in  \cite{Foll1}, it is well known
that for any $ u \in {\mathcal S}(\Ra^{d})$ one has
\begin{eqnarray}
 u  = (2 \pi  \hbar)^{-d} \int_{\Ra^{2d}}  \left\langle \varphi_{\hbar}^a  , u
\right\rangle_{L^2} \varphi_{\hbar}^a \ \mbox{d}a
\end{eqnarray}
 where $ \varphi_{\hbar}^a $ is  defined by $ \reff{notationetats}$ with $ \mu
=  1/2$.  This implies  in  particular that,  for  any  $ \tilde{\varphi}  \in
{\mathcal S} (\Ra^d) $,
\begin{eqnarray}
 || u   ||_{L^2}^2   ||  \tilde{\varphi}||_{L^2}^2   =   (2  \pi   \hbar)^{-d}
\int_{\Ra^{2d}}    \left|   \left\langle    \tilde{\varphi}_{\hbar}^a    ,   u
\right\rangle_{L^2} \right|^2 \ \mbox{d}a .
\end{eqnarray}
This decomposition on $L^2(\Ra^d)$, known as the coherent states decomposition
especially when $  \varphi(q)= \eta (q) = \pi^{-d/4}  e^{-q^2/2} $, gives rise
to a decomposition on $ {\mathcal H}_{\hbar}(\kappa) $
\begin{eqnarray}
S_{\hbar}(\kappa)  u   =  (2  \pi   \hbar)^{-d}  \int_{\Ta^{2d}}  \left\langle
\varphi^a_{\hbar,\kappa}   ,   S_{\hbar}(\kappa)  u   \right\rangle_{{\mathcal
H}_{\hbar}(\kappa)} \varphi^a_{\hbar,\kappa} \ \mbox{d}a, \label{decomptore}
\end{eqnarray}
with   the  notation   of   $   \reff{projection}  $.   This   is
proven   in \cite{BoDB}.  Note  the important  consequence  of
that formula: for  any  $ \tilde{\varphi} \in {\mathcal S}
(\Ra^{d}) $
\begin{eqnarray}
(2  \pi  \hbar)^{-d}  \int_{\Ta^{2d}}  \left|  \left\langle  S_{\hbar}(\kappa)
\tilde{\varphi}_{\hbar}^a,  S_{\hbar}(\kappa)   u  \right\rangle  \right|^2  \
\mbox{d}a  =  C_{\tilde{\varphi}} \left|  \left|  S_{\hbar}(\kappa) u  \right|
\right|_{{\mathcal H}_{\hbar}(\kappa)}^2  , \qquad  \forall \ u  \in {\mathcal
S}(\Ra^d). \label{controlnorm}
\end{eqnarray}
These  decompositions  are  particularly  convenient since  one  knows  rather
precisely the action of pseudodifferential operators
on functions of the form $ \reff{notationetats}  $, as we shall see in Lemma $
\refe{below} $ below.  Motivated by Lemma $ \refe{spectralcutoff}  $, we shall
consider functions $ f $ depending possibly  on $ \hbar $. Let $ \varepsilon >
0 $ and assume that $ r_{\hbar} $ is a sequence such that
$$ r_{\hbar}  \geq \hbar^{1/2  - \varepsilon} $$  and let  $ f_{\hbar} $  be a
family of functions in $ {\mathcal B}(\Ra^{2d}) $ such that
\begin{eqnarray}
 \left|    \partial^{\gamma}     f^{\hbar}(x)    \right|    \leq    C_{\gamma}
r_{\hbar}^{-|\gamma|}, \qquad x \in \Ra^{2d} . \label{dilatesymbole}
\end{eqnarray}
\begin{lemm} \label{below} There exists a family $ P_{\gamma} $ of differential operators
 with polynomial coefficients  (independent of $ \hbar $) such  that for any $
f^{\hbar}  $  as  above   and  any  $  M  >  0  $,   there  exists  symbols  $
f^{(\hbar,M,\gamma)}  $  satisfying  $  \reff{dilatesymbole}  $  as  well  and
differential  operators  $   Q^{M}_{\gamma}  $  with  polynomial  coefficients
(independent of $ \hbar $ too) such that
$$ O  \! p^W  (f^{\hbar}) U_{\hbar}(a) \varphi_{\hbar}  = \sum_{|\gamma|  < M}
\hbar^{|\gamma|/2}  \partial^{\gamma} f^{\hbar}(a)  U_{\hbar}  (a) (P_{\gamma}
\varphi)_{\hbar} +  \hbar^{M \varepsilon}  \sum_{|\gamma| \leq 2  M} O  \! p^W
(f^{\hbar,M,\gamma})  U_{\hbar}(a)   (  Q^M_{\gamma}\varphi  )_{\hbar}   .  $$
Whenever  $  A  =  P_{\gamma}  $  or  $ Q_{\gamma}^M  $,  we  have  set  $  (A
\varphi)_{\hbar} (q) = h^{-d/4} (A \varphi)(q / \hbar^{1/2} ) $.
\end{lemm}

\noindent {\it Proof.} It is essentially standard. Since $
U_{\hbar}(-a) O  \!p^{W}(f) U_{\hbar}(a) = O \!p^W  (f(. + a)) $,  we are left
with the case $ a = 0 $. Then, the result simply follows by writing the Taylor
expansion of $ f^{\hbar} $ at $ 0 $ and integrating by parts. \finpreuve

\medskip

\noindent {\it Remark.} The operators $ P_{\gamma} $ can be
computed explicitly and in particular $ P_0 = I $.

\medskip

Combining this result and $ \reff{Laformule}  $, it is not hard to deduce that
for   any    $   f^{\hbar}    \in   C^{\infty}(\Ta^{2d})   $    satisfying   $
\reff{dilatesymbole} $, one has, for all $ M > 0 $,
\begin{eqnarray}
\left| \left| O  \!p^W (f^{\hbar}) \varphi^{a}_{\hbar,\kappa} - \sum_{|\gamma|
<  M}   \hbar^{|\gamma|/2}  \partial^{\gamma}  f^{\hbar}(a)  S_{\hbar}(\kappa)
U_{\hbar}(a)    (P_{\gamma}   \varphi)_{\hbar}    \right|   \right|_{{\mathcal
H}_{\hbar}(\kappa)} \leq C \hbar^{M \varepsilon} \label{expansionantiwick}
\end{eqnarray}
uniformly with respect  to $ a \in [ 0  ,1 )^{2d} $. We are  now ready for the
proof of Lemma $ \refe{spectralcutoff} $.

\medskip

\noindent {\bf Proof of Lemma ${\bf \refe{spectralcutoff} }$.} We
only have  to show the existence of  a sequence $ r_{\hbar}  \geq \hbar^{1/2 -
\varepsilon}  $  for  some  $  \varepsilon  >  0  $,  satisfying  $  r_{\hbar}
\rightarrow 0 $, such that, if $ 0  \leq \chi \leq 1 $ is supported close to $
0 $ and $ \equiv 1 $ near $ 0 $ then
$$ \chi_{\hbar}(a) := \sum_{n  \in \Za^{2d}} \chi \left( \frac{a+n}{r_{\hbar}}
\right) $$ will satisfy  the result. Let us fix $ \varepsilon  > 0 $. Then for
any  sequence  $  r_{\hbar}  \geq  \hbar^{1/2  -  \varepsilon}  $,  using  the
Proposition $ \refe{fcomp} $, one has
$$  O   \!p^W  (1-\chi_{\hbar})^2   =  \sum_{j  <   M}  \hbar^{j}  O   \!  p^W
(\chi_{j,\hbar}) + o (1) $$ in operator norm, provided $ M = M (\varepsilon) $
is  large   enough.  The   symbols  $  \chi_{j,\hbar}   $  are  such   that  $
\partial^{\gamma} \chi_{j,\hbar} = {\mathcal O}(r_{\hbar}^{-|\gamma|-2j})$ and
$ \chi_{0,\hbar} = (1- \chi_{\hbar})^2  $, thus using $ \reff{decomptore} $, $
\reff{controlnorm} $ and $ \reff{expansionantiwick} $, one has
$$  \left| \left|  O \!  p^W  (1 -  \chi_{\hbar}) \Psi_{\hbar,\kappa}  \right|
\right|^2 = (2 \pi  \hbar)^{-d} \int_{\Ta^{2d}} (1 - \chi_{\hbar}(a))^2 \left|
\left\langle   \eta^a_{\hbar,\kappa}   ,   \Psi_{\hbar,\kappa}   \right\rangle
\right|^2 \ \mbox{d}a + o (1)
$$ using  also the fact that $  || \Psi_{\hbar,\kappa} || \rightarrow  1 $. By
Taylor    formula,   there    exists   a    function   $    \tilde{\chi}   \in
C^{\infty}(\Ta^{2d}) $, independent of $  \hbar $, such that $ \tilde{\chi}(0)
=  0  $  and  $  (1-\chi_{\hbar}(a))^2 \leq  \tilde{\chi}^2(a)  /  r_{\hbar}^2
$. Since
\begin{eqnarray} (2
\pi   \hbar)^{-d}   \int_{\Ta^{2d}}   \tilde{\chi}(a)^2  \left|   \left\langle
\eta^a_{\hbar,\kappa}   ,   \Psi_{\hbar,\kappa}   \right\rangle  \right|^2   \
\mbox{d}a \rightarrow 0 \label{slow}
\end{eqnarray}
by $ \reff{concentrationzero} $ applied to $ f = \tilde{\chi}^2 $, we see that
$  || O  \! p^W  (1  - \chi_{\hbar})  \Psi_{\hbar,\kappa} ||  \rightarrow 0  $
provided $ r_{\hbar}^2 \rightarrow 0 $  more slowly than the left hand side of
$ \reff{slow}  $. Furthermore  there is no  restriction to choose  $ r_{\hbar}
\geq \hbar^{1/2 - \varepsilon} $. Finally, we remark that
$$  O   \!  p^W  (\chi_{\hbar})  \Psi_{\hbar,\kappa}  =   (2  \pi  \hbar)^{-d}
\int_{\Ta^{2d}}   \left\langle  \eta^a_{\hbar,\kappa}   ,
\Psi_{\hbar,\kappa} \right\rangle \chi_{\hbar}(a)
\eta^a_{\hbar,\kappa} \  \mbox{d}a + o (1) $$ by $
\reff{decomptore} $, $ \reff{controlnorm} $ and $
\reff{expansionantiwick} $ again which completes the proof of $
\reff{speed} $. For the converse, we note that
$$  \left\langle   \Psi_{\hbar,\kappa},   O   \!  p^W   (f)   \Psi_{\hbar,\kappa}
\right\rangle  - (2 \pi \hbar)^{-d} \int_{\Ta^{2d}}
\chi_{\hbar}(a)  \left\langle \eta^a_{\hbar,\kappa}   ,
\Psi_{\hbar,\kappa} \right\rangle \left\langle
\Psi_{\hbar,\kappa} , O \! p^W (f) \eta^a_{\hbar,\kappa}
\right\rangle
 \ \mbox{d}a \rightarrow 0.
$$
The result follows then easily from the dominated convergence
theorem using $ \reff{expansionantiwick} $ and $
\reff{controlnorm} $. \finpreuve

\end{document}